\documentclass[twocolumn]{aastex631}

\usepackage{amsmath}
\usepackage{color}
\usepackage{rotating} 

\usepackage[mathlines]{lineno}
\begin{document}
\title{Beam Measurements of Full Stokes Parameters for the FAST L-band 19-beam Receiver}

\correspondingauthor{Xunzhou Chen}
\email{cxzchen123@gmail.com}

\author[0000-0003-3957-9067]{Xunzhou Chen}
\affiliation{School of Science, Hangzhou Dianzi University, Hangzhou, People’s Republic of China}
\affiliation{Research Center for Astronomical Computing, Zhejiang Laboratory, Hangzhou 311100, People’s Republic of China}

\author[0000-0001-8516-2532]{Tao-Chung Ching}
\altaffiliation{Tao-Chung Ching is a Jansky Fellow of the National Radio Astronomy Observatory.}
\affiliation{National Radio Astronomy Observatory, 1003 Lopezville Road, Socorro, NM 87801, USA}

\author{Di Li}
\affiliation{Department of Astronomy, Tsinghua University, Haidian District, Beijing, 100084, P. R. China}
\affiliation{National Astronomical Observatories, Chinese Academy of Sciences, Beijing 100101, People’s Republic of China}

\author{Carl Heiles}
\affiliation{University of California, Berkeley, CA 94720, USA}

\author[0000-0002-4217-5138]{Timothy Robishaw}
\affiliation{Dominion Radio Astrophysical Observatory, Herzberg Astronomy \& Astrophysics Research Centre, National Research Council Canada, P.O.\ Box 248, Penticton, BC, V0H 1K0, Canada}

\author{Xuan Du}
\affiliation{Dominion Radio Astrophysical Observatory, Herzberg Astronomy \& Astrophysics Research Centre, National Research Council Canada, P.O.\ Box 248, Penticton, BC, V0H 1K0, Canada}

\author{Marko Krco}
\affiliation{National Astronomical Observatories, Chinese Academy of Sciences, Beijing 100101, People’s Republic of China}

\author{Peng Jiang}
\affiliation{National Astronomical Observatories, Chinese Academy of Sciences, Beijing 100101, People’s Republic of China}
\affiliation{CAS Key Laboratory of FAST, National Astronomical Observatories, Chinese Academy of Sciences, Beijing 100101, People’s Republic of China}

\author{Qingliang Yang}
\affiliation{National Astronomical Observatories, Chinese Academy of Sciences, Beijing 100101, People’s Republic of China}

\author{Jiguang Lu}
\affiliation{National Astronomical Observatories, Chinese Academy of Sciences, Beijing 100101, People’s Republic of China}

\begin{abstract}
The Five-hundred-meter Aperture Spherical radio Telescope (FAST) has been fully operational since 11 January 2020. We present a comprehensive analysis of the beam structure for each of the 19 feed horns on FAST’s L-band receiver across the Stokes $I$, $Q$, $U$, and $V$ parameters. Using an on-the-fly mapping pattern, we conducted simultaneous sky mapping using all 19 beams directed towards polarization calibrators J1407+2827 and J0854+2006 from 2020 to 2022. Electromagnetic simulations were also performed to model the telescope’s beam patterns in all Stokes parameters. Our findings reveal a symmetrical Gaussian pattern in the Stokes $I$ parameter of the central beam without strong sidelobes, while the off-center beams exhibit significant asymmetrical shapes that can be fitted using a combination of log-normal and Gaussian distributions. The inner beams have higher relative beam efficiencies and smaller beam sizes compared to those of the outer beams. The sidelobes of the inner beams contribute approximately 2\% of the total flux in the main lobe, increasing to 5\% for outer beams, with a peak at 6.8\%. In Stokes $U$, a distinct four-lobed cloverleaf beam squash structure is observed, with similar intensity levels in both inner and outer beams. In Stokes $V$, a two-lobed beam squint structure is observed in the central beam, along with a secondary eight-lobed structure. The highest squint peak in Stokes $V$ is about 0.3\% of the Stokes $I$ in the outer beams. These results align closely with the simulations, providing valuable insights for the design of radio multi-beam observations.
\end{abstract}


\section{Introduction} \label{sec:intro}

Numerous large radio experiments have been established in recent years, with a significant goal being the mapping of neutral hydrogen (H\uppercase\expandafter{\romannumeral1}) intensity in the Universe through its 21 cm emission line. The Five-hundred-meter Aperture Spherical radio Telescope (FAST), with its 500-meter diameter, is the most sensitive single-dish telescope operating in the frequency range of 1050 to 1450 MHz (L-band). It was officially commissioned and accepted for national use on January 11, 2020. The commissioning phase of FAST was initiated following the completion of its construction on September 25, 2016, with detailed insights available in \citet{jiang2019}. FAST's 19-beam L-band receiver enables simultaneous observations across 19 distinct beams, enhancing survey speed and spatial coverage. The Commensal Radio Astronomy FasT Survey (CRAFTS; \citealt{Li2018}) is a large-scale observational effort with the primary goal of obtaining data simultaneously for Galactic H\uppercase\expandafter{\romannumeral1} imaging and the detection of H\uppercase\expandafter{\romannumeral1} galaxies using multiple backends.

Accurate characterization of beam shapes is essential; for instance, a 1\% error in beam size can introduce a 4\% systematic uncertainty in the power spectrum amplitude for H\uppercase\expandafter{\romannumeral1} intensity mapping \citep{Chang2015}. Consequently, recent studies have focused on the beam shapes of FAST in Stokes parameters, underscoring their significance for achieving precise polarization measurements and minimizing systematic errors in astrophysical observations. \citet{jiang2020} found that the morphology of the sidelobes and their angular distances from the beam center depend on the observation frequency. They also noted that while the main beam’s noncircular morphology and coma---a distortion pattern causing asymmetrical, tail-like elongation in a telescope’s beam---are negligible for the central beam, these effects become increasingly pronounced for the outer beams as their angular distance from the center increases. \citet{Sun2021} performed 2-D elliptical Gaussian fittings on total-intensity maps for each beam at each frequency channel, finding minimal differences between the major and minor beam widths. \citet{Jing2024} also modeled the beam shapes as 2-D Gaussian functions, corroborating the findings of \citet{jiang2020} by showing that the off-center beams exhibit irregular beam shapes with minor size variations. In this study, we aim to comprehensively characterize the full-Stokes beam patterns of the 19-beam receiver on FAST from 2020 to 2022, establishing a long-term record beneficial for future improvements and interpretation of observational results. The analysis methods developed in this study can also be applied to characterize the beam structure of multi-beam receivers on other radio telescopes. 


We have been conducting a long-term study of the polarization performance of the FAST telescope since full operation commenced in 2020. The calibration of the on-axis Mueller matrices of the 19-beam receiver with the narrow-band back-end within the full illumination of the telescope (zenith angle $\leq 26.4^{\circ}$) will be presented in Ching et al. (submitted, arXiv:2411.18763). This work focuses on the beam structures of the full Stokes parameters to present the off-axis polarization performance of the 19-beam receiver. Calibration of beam patterns in radio telescopes typically involves observing bright astronomical radio continuum sources such as the sun (e.g., \citealt{Kraus1966}), the moon \citep{Tello2013}, and known bright radio sources like Cassiopeia A, Taurus A, Cygnus A, and Virgo A \citep{Baars1977}. By drift-scanning these sources over the beam extent, the beam shape convolved with the source can be measured. This study employs a multi-beam on-the-fly (MultiBeamOTF) mapping mode for simultaneous sky mapping with 19 beams targeting the quasar J1407+2827 (Mrk 668) and the blazar J0854+2006 (OHIO J 287). We characterize the size, shape, and efficiency of FAST’s 19 beams in Stokes $I$ near 1420.4 MHz, examining their variations from 2020 to 2022, along with beam shapes in Stokes $Q$, $U$, and $V$. An electromagnetic (EM) simulation was also carried out to model the beam structure in all four Stokes parameters. Section \ref{sec2} details the methods for measuring beam shape, including observation, data reduction, and EM simulation. Results on beam shapes and sidelobes are presented in Section \ref{res}, with a summary in Section \ref{con}.

\begin{figure*}
   \centering
   \includegraphics[width=0.9\textwidth, angle=0]{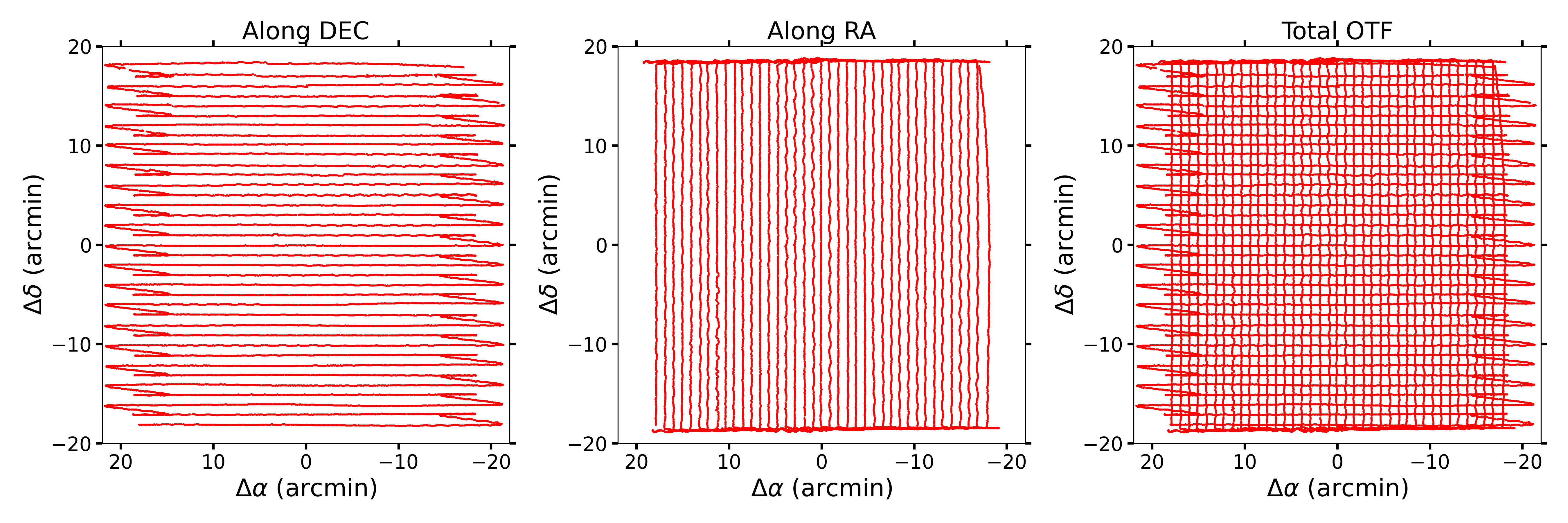}
   \caption{\textbf{Left:} The telescope track during a raster scan along the Dec direction. \textbf{Middle:} The telescope track during a raster scan along the RA direction. \textbf{Right:} The overall telescope track during a raster scan. The $x$ and $y$ axes represent the RA and Dec offsets relative to the equatorial coordinates of J0854+2006 observed in 2021.}
   \label{Fig1}
\end{figure*}
   
\section{Observations, Data Reduction, and Simulation}\label{sec2}
\subsection{Observation}
\label{sect:Obs}

To determine the beam structures, their dependence on zenith angle, and time stability, multiple observations were conducted over different sources and epochs. 
We chose the VLA polarization calibrators\footnote{https://science.nrao.edu/facilities/vla/docs/manuals/obsguide \allowbreak /modes/pol} J0854+2006 and J1407+2827 as our sources. Among the VLA polarization calibrators, J0854+2006 is the one closest to the zenith of FAST. OTF observations of J0854+2006, each lasting approximately 1.7 hours per day, were carried out on 2020/09/10, 2020/09/11, 2021/10/06, and 2021/10/07. J1407+2827 is known for low polarization at L-band and hence is an ideal target to map the Stokes $Q$, $U$, and $V$ beams.
OTF observations of J1407+2827, each lasting approximately 2.1 hours per day, were carried out on 2022/09/19 and 2022/09/21. Raster scans around the transit of the targets using the MultiBeamOTF mode of the telescope with a scan spacing of 1 arcmin and a scanning speed of 15 arcsec/sec were performed in either the declination (Dec) or right ascension (RA) direction, covering a sky region of approximately $20^\prime \times 20^\prime$, with a consistently low zenith angle ($\leq 18.7^\circ$) for all observations. 
Offsets in RA and Dec relative to the equatorial coordinates of all sources were computed according to \S~2.2.2 of \citet{jiang2020}. The full-polarization correlation products of the signals were simultaneously recorded using the ROACH backend\footnote{https://casper.berkeley.edu/wiki/ROACH-2\_Revision\_2}, featuring 65,536 spectral channels in each polarization. The spectral bandwidth, centered at the frequency of the 21-cm H\uppercase\expandafter{\romannumeral1} line, was 31.25 MHz, with a channel spacing of $\sim$477 Hz. Data acquisition employed a sampling rate of 0.1~s and noise diode injection cadence of 0.2~s/1.8~s for on/off switching. During raster scan observations, the feed’s phase center is measured in a real-time measurement system (see \citealt{jiang2019} for details) and converted into the telescope’s central beam (M01) pointing in horizontal coordinates (alt, az). The corresponding equatorial coordinates (RA, Dec in J2000) are then derived, accounting for precession, nutation, aberration, and atmospheric refraction. Figure \ref{Fig1} illustrates an example of the central beam track for a raster scan targeting J0854+2006 in 2021.

\subsection{Data reduction}
We used the IDL RHSTK package\footnote{http://w.astro.berkeley.edu/heiles/} developed by C.~Heiles and T.~Robishaw to reduce the FAST raw data and generate the Stokes $I$, $Q$, $U$, and $V$ spectra. Widely adopted for Arecibo and GBT polarization data reduction, this package excels in executing gain and phase calibrations for the two polarization paths, bandpass calibrations for the four correlated spectra, and polarization calibrations. 
The process of polarization calibration used for the data in this work was provided in Ching et al. (submitted, arXiv:2411.18763).
Based on the Stokes $I$, $Q$, $U$, and $V$ spectra calibrated using the RHSTK package, the Asymmetrically Reweighted Penalized Least Squares smoothing algorithm, as proposed by \citet{Baek2015}, was employed to eliminate the baselines of the spectra. With a pixel size of $0.5^\prime$, beam maps were generated using a Gaussian smoothing kernel with a full width at half maximum (FWHM) of $7.9^{\prime\prime}$ to account for the uncertainty in the pointing accuracy of FAST \citep{jiang2020} and to convert the irregular sampling into a regular grid. 

\subsection{Simulated beam patterns}

The Stokes $I$, $Q$, $U$, and $V$ beam patterns of the FAST telescope were modeled using the method described in \citet{Du2022}. First, the radio telescope was simulated as a transmitting antenna using the commercial software \citet{CST} and \citet{GRASP}. The far-field radiation pattern of the feed horn was simulated using CST. Only one feed horn (instead of a 19-horn array) was simulated, as mutual-coupling between the horns is minimal \citep{smith2017}. The reflector antenna was then simulated in GRASP by placing the simulated feed pattern at the corresponding positions of the 19 horns on the receiver, one at a time. The results are two sets of 19 radiation patterns of the telescope, one for each linear polarization, in the form of far-field electric fields. For each of the 19 beams, the 2~\texttimes~2 Jones matrix of the telescope (in transmitting mode) can be constructed using these simulated far-field electric fields. The Jones matrix in the receiving mode was then deduced from the transmitting-mode Jones matrix using the theorem of reciprocity \citep{Potton2004}, and further converted to the 4~\texttimes~4 Mueller matrix. For each of the 19 feed horns, 16 beam patterns are produced---one for each of the Mueller matrix elements---that fully characterize the telescope's response to unpolarized, linearly-polarized, and circularly-polarized incoming waves. Among them, the Stokes $I$, $Q$, $U$, and $V$ beam patterns presented in this paper describe the telescope's response to unpolarized incoming light.

\section{Result}\label{res}
\subsection{Fitting to the Stokes $I$}
\subsubsection{The central beam}

  \begin{figure*}
   \centering
   \includegraphics[width=1.0\textwidth, angle=0]{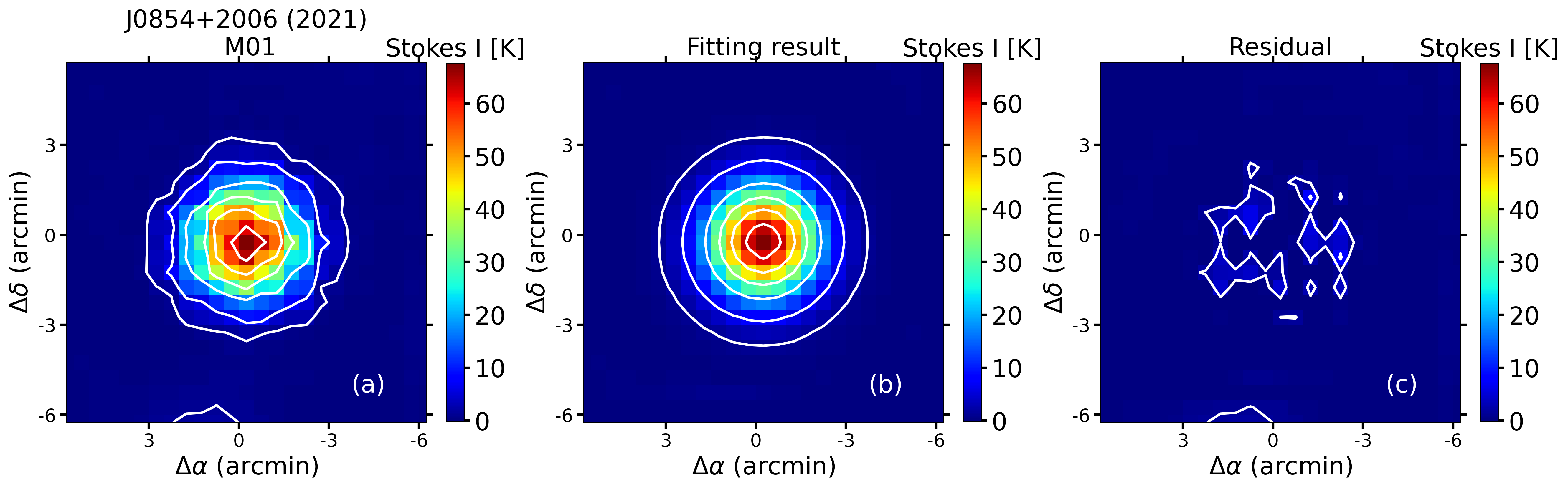}
   \caption{\textbf{(a)} Processed Stokes $I$ data for the central beam, mapping the blazar J0854+2006 during observations conducted in 2021. The feature near the bottom of the map is attributed to a nearby secondary source, J0854+1959, which is close enough to contribute to the observed data. \textbf{(b)} Results of Gaussian fitting applied to the Stokes $I$ data. \textbf{(c)} Residual map showing the differences between the Stokes $I$ data and the Gaussian fitting results. White solid lines indicate contour levels at 2\%, 10\%, 30\%, 50\%, 70\%, and 90\% of the peak flux. The Stokes $I$ power values are expressed in units of kelvin (hereafter referred to as K).}
   \label{Fig2}
   \end{figure*}

  \begin{figure*}
   \centering
   \includegraphics[width=\textwidth, angle=0]{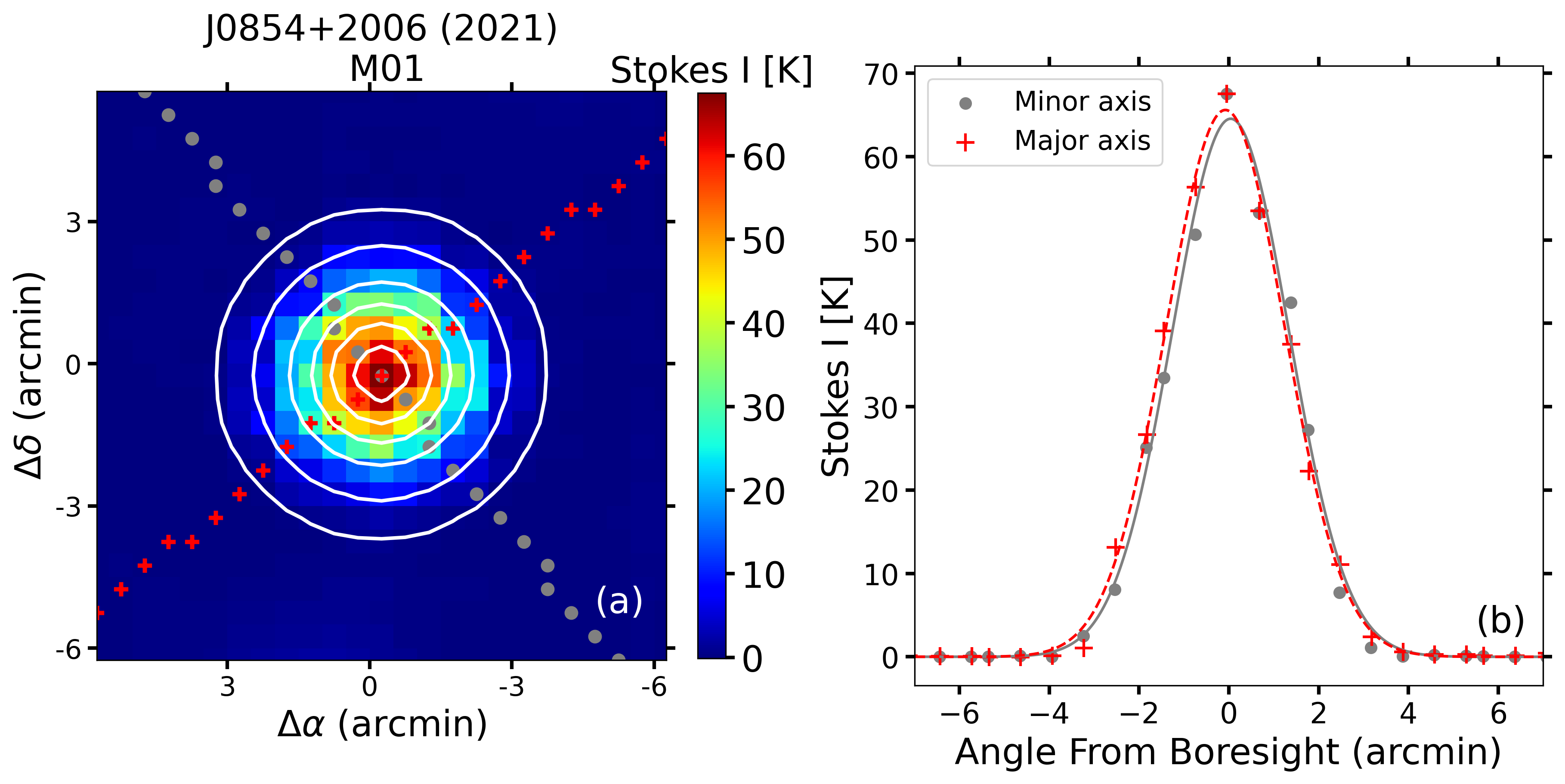}
   \caption{\textbf{(a)} Gaussian fitting results with cuts along the major and minor axes for the central beam. Red plus signs and gray circles represent the nearest pixels to the straight lines passing through the major and minor axes, respectively. Contours (white solid lines) show intensity levels at 2\%, 10\%, 30\%, 50\%, 70\%, and 90\% of the peak flux. \textbf{(b)} Beam power data through a single antenna beam pattern. The cut along the major axis is represented by red plus signs with the corresponding Gaussian fit (red dashed line), while the minor axis cut is shown by gray circles with its Gaussian fit (gray solid line). The beam cuts are fairly symmetrical, consistent with the 2-D symmetry of the central beam.}
   \label{Fig3}
   \end{figure*}

Figure \ref{Fig2}(a) illustrates the symmetrical Stokes $I$ profile of the central beam. To model the convolved beam pattern, a 2-D Gaussian fitting technique is applied. Notably, the nearby weak radio source J0854+1959, located to the southeast of J0854+2006, contributes a blob. Since it does not blend into the beam structure as measured on J0854+2006, it has been excluded from all following fitting and analysis. The beam’s coordinate transformation and power distribution are described by the following equations:
\begin{equation}\label{eq1}
X = x_0\cos\phi + y_0\sin\phi,\ Y = y_0\cos\phi - x_0\sin\phi \, ,
\end{equation}
%
\begin{multline}\label{eq2}
P(X,Y)=\frac{A}{2\pi\sigma_1\sigma_2} \\
\cdot \exp\left\{-\frac{1}{2}\left[\frac{(X-\mu_1)^2}{{\sigma_1}^2}+\frac{(Y-\mu_2)^2}{{\sigma_2}^2}\right]\right\} \, .
\end{multline}
%
In these equations, $x_0$ and $y_0$ represent the initial Cartesian coordinates of the beam pattern. The variable $\phi$ denotes the position angle, which rotates the coordinate system by $\phi$ degrees to align with the major axis of the Gaussian distribution. It is defined as the angle of the major axis measured from North towards East. $P(X,Y)$ represents the power distribution of the beam at the new coordinates $X$ and $Y$. The term $A$ is a constant related to the intensity of the beam, and $(\mu_1, \mu_2)$ are the coordinates of the beam center. The parameters $\sigma_1$ and $\sigma_2$ are the standard deviations along the major and minor axes, respectively. Figure \ref{Fig2}(b) presents the 2-D Gaussian fitting result for the central beam of Stokes $I$, showing a symmetrical profile. The deviation between the processed data and the Gaussian fitting result is minimal, as shown in Figure \ref{Fig2}(c). Additionally, Figure \ref{Fig3} shows the cuts along the major and minor axes, which follow Gaussian profiles along both axes. This 2-D Gaussian modeling confirms the consistent shape of the central beam over three years, with no detectable sidelobes. The well-behaved central beam, free from significant sidelobes, is crucial for ensuring the integrity of collected observational data.

\subsubsection{The off-center beams}
\begin{figure*}
   \centering
   \includegraphics[width=1.0\textwidth, angle=0]{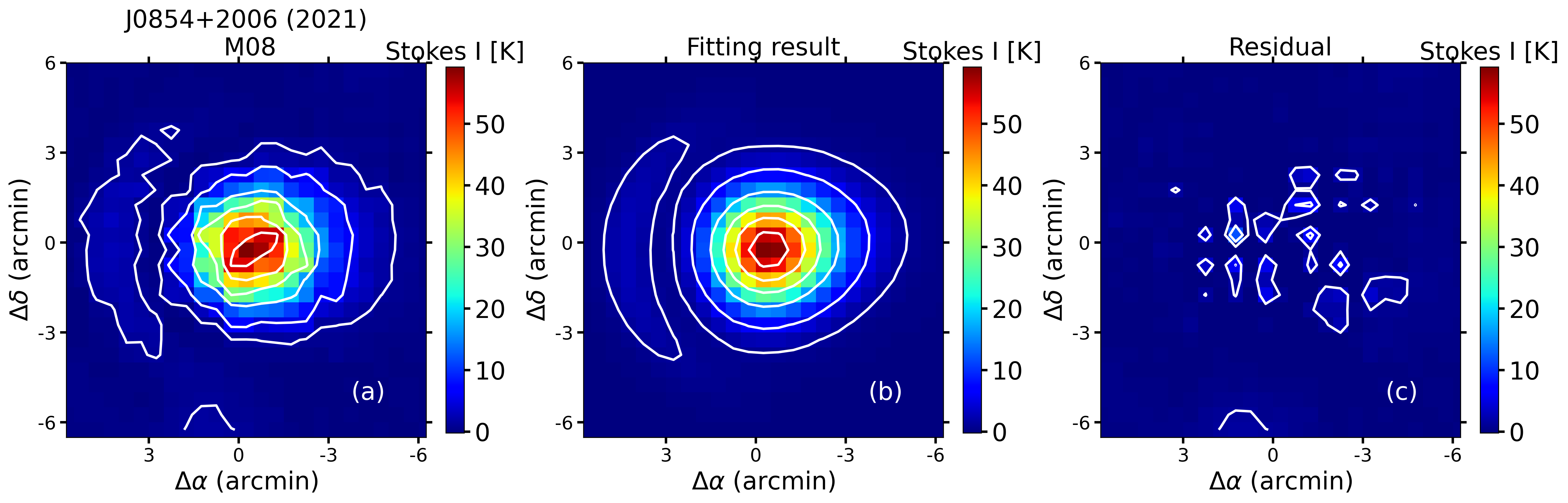}
   \caption{\textbf{(a)} Processed Stokes $I$ data for beam M08, mapping the blazar J0854+2006 during observations conducted in 2021. The feature near the bottom of the map is attributed to a nearby secondary source, J0854+1959, which is close enough to contribute to the observed data. \textbf{(b)} Results of the log-normal combined with Gaussian fitting applied to the Stokes $I$ data. \textbf{(c)} Residual map showing the differences between the Stokes $I$ data and the fitting results. White solid lines indicate contour levels at 2\%, 10\%, 30\%, 50\%, 70\%, and 90\% of the peak flux.}
   \label{Fig4}
   \end{figure*}

As noted in \citet{jiang2020} and \citet{Jing2024}, FAST is equipped with a 19-beam receiver system. While the central beam exhibits a consistent and well-defined shape, the off-center beams display more complex and irregular structures. The shapes of the off-center beams are comatic, with slight differences in beam size. The differences between the central beam and the off-center beams, including variations in their sidelobes, are from the lateral displacement of each feed horn from the focus of the telescope. For instance, as depicted in Figure \ref{Fig4}(a), the response from beam M08 (according to labels of Figure \ref{Fig6}) showcases a notably intricate distribution. Within beam M08, two additional asymmetrical structures become apparent: 
\begin{enumerate}
    \item[1] Due to coma, an off-axis point appears as a blurred, comet-like shape with a distinct head and tail, hence the term \textit{coma} \citep{Strong1958,Guenther1990,Pedrotti1993}. The coma spans the region from $\Delta \alpha \sim-6^\prime$ to $-3^\prime$ and $\Delta \delta \sim -3^\prime$ to $3^\prime$, attributed to the lateral offset of the horn from the focus of the reflecting surface. Coma is also observed in other single-dish radio telescopes with multi-beam receivers, such as Arecibo and Parkes \citep{Staveley1996,heiles2001,Giovanelli2005}. The orientation of the major axis of the elongated main beam is aligned with feed horn's offset from the focus.
    \item[2] A crescent-like sidelobe (also frequently refered to as a ``coma lobe''; \citealt{Stutzman1981,Milligan2005}) emerging at $\Delta \alpha \sim3^\prime$ to $5^\prime$ and $\Delta \delta \sim-4^\prime$ to $4^\prime$. The sidelobe is symmetric across the major axis of the main beam and is situated on the far side of the telescope focus.
\end{enumerate}
Consequently, conventional 2-D Gaussian fitting methods become inadequate necessitating the adoption of a novel technique for characterizing the sidelobe. In this context, we employ a log-normal distribution combined with a Gaussian distribution to model the main lobe for beam M08. This approach introduces an additional parameter, $a$, compared to Equation \ref{eq2}, while retaining the other parameters unchanged:
\begin{equation}\label{eq3}
X_1 = \ln(X+a),\ Y_1 = Y
\end{equation}
The power distribution can then be described by $P(X_1,Y_1)$. The parameter $a$ is related to the beam center and the FWHM along the elongated major axis of the main beam.

For the sidelobe, akin to Equation \ref{eq2}, we also construct a 2-D Gaussian distribution, yet the fitting variables are modified to $r$ and $\phi$. Notably, sidelobes appear predominantly on one side of the main beam, at an orientation opposite to the focus. Consequently, we fit only the segment containing the sidelobe. Therefore, the fitting region for the sidelobe is restricted to $X_0 \leq 0^\prime$:
\begin{equation}\label{eq4}
X = x_0\cos\phi_c+y_0\sin\phi_c,\ Y = y_0\cos\phi_c-x_0\sin\phi_c \, ,
\end{equation}
\begin{equation}\label{eq5}
r = \sqrt{X^2+Y^2},\ \phi = \arctan 
\left( \frac{Y}{X} \right) \, ,
\end{equation}
\begin{equation}\label{eq6}
p(r,\phi) = \frac{A_1}{2\pi\sigma_3\sigma_4} \cdot \exp\left\{-\frac{1}{2}\left[\frac{(r-r_c)^2}{{\sigma_3}^2}+\frac{(\phi)^2}{{\sigma_4}^2}\right]\right\} \, .
\end{equation}
Here, $x_0$ and $y_0$ represent the initial Cartesian coordinates of the beam pattern, $A_1$ represents a constant, $\phi_c$ denotes the position angle of the sidelobe, ($r_c, \phi_c$) represents the center position corresponding to the peak flux, and $\sigma_3$ and $\sigma_4$ represent standard deviations related to the width and field angle of the sidelobe, respectively.

\begin{figure*}
   \centering
   \includegraphics[width=\textwidth, angle=0]{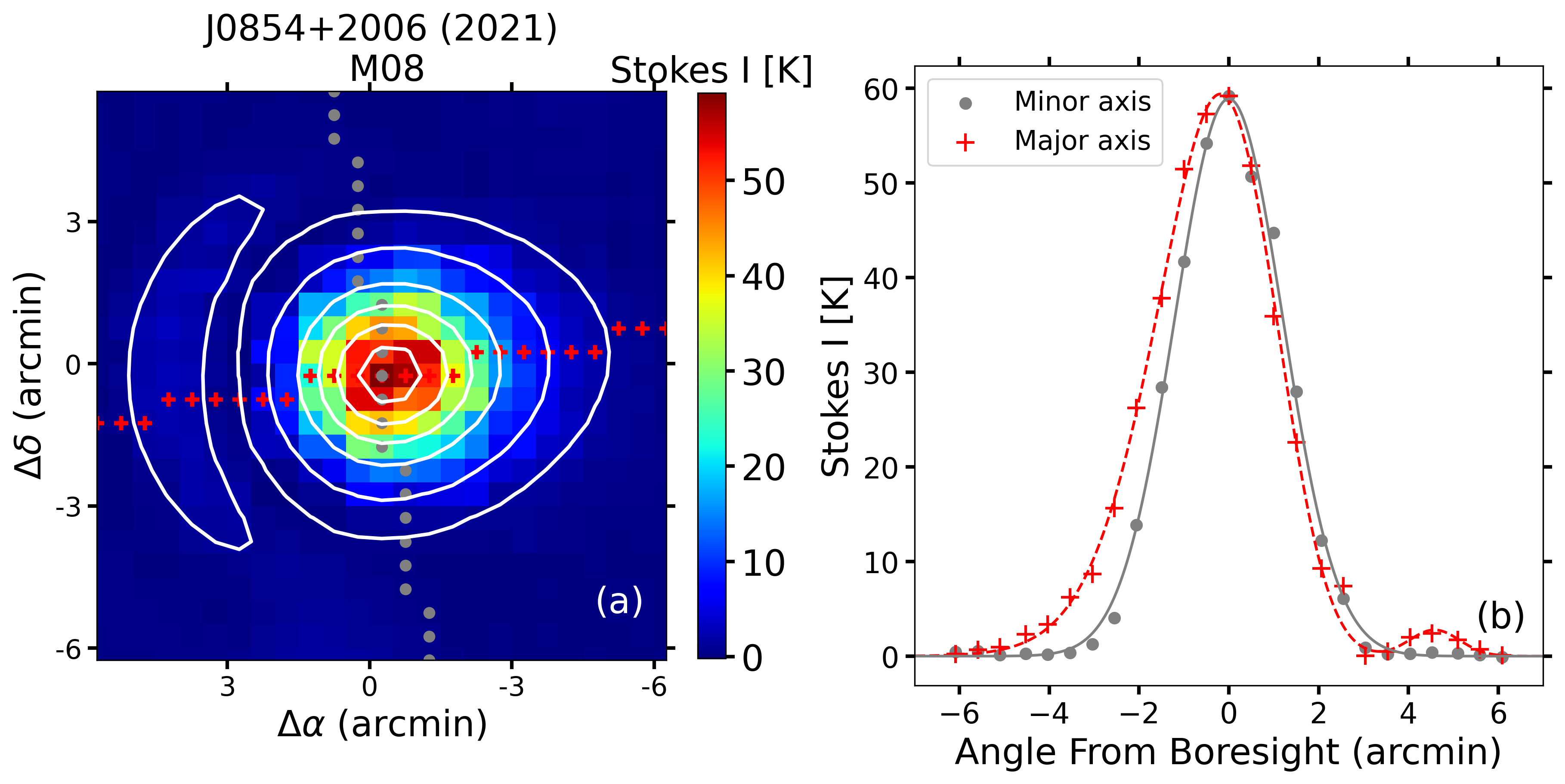}
   \caption{\textbf{(a)} The log-normal combined with Gaussian fitting results with cuts along the major and minor axes for beam M08. The sidelobes are also well-fitted by a Gaussian distribution. Red plus signs and gray circles represent the nearest pixels to the straight lines passing through the major and minor axes, respectively. Contours (white solid lines) show intensity levels at 2\%, 10\%, 30\%, 50\%, 70\%, and 90\% of the peak flux. \textbf{(b)} Beam power data through a single antenna beam pattern. The cut along the major axis is represented by red plus signs. The red dashed line shows the fitting results including the log-normal fitting for the main lobe and the Gaussian fitting for the sidelobe. The cut along the minor axis is depicted by gray circles with its Gaussian fit (gray solid line). The vertical beam cut is fairly symmetrical and the horizontal cut is asymmetrical due to the sidelobe.
   }
   \label{Fig5}
   \end{figure*}

\begin{figure*}
\centering
\includegraphics[width=0.9\textwidth, angle=0]{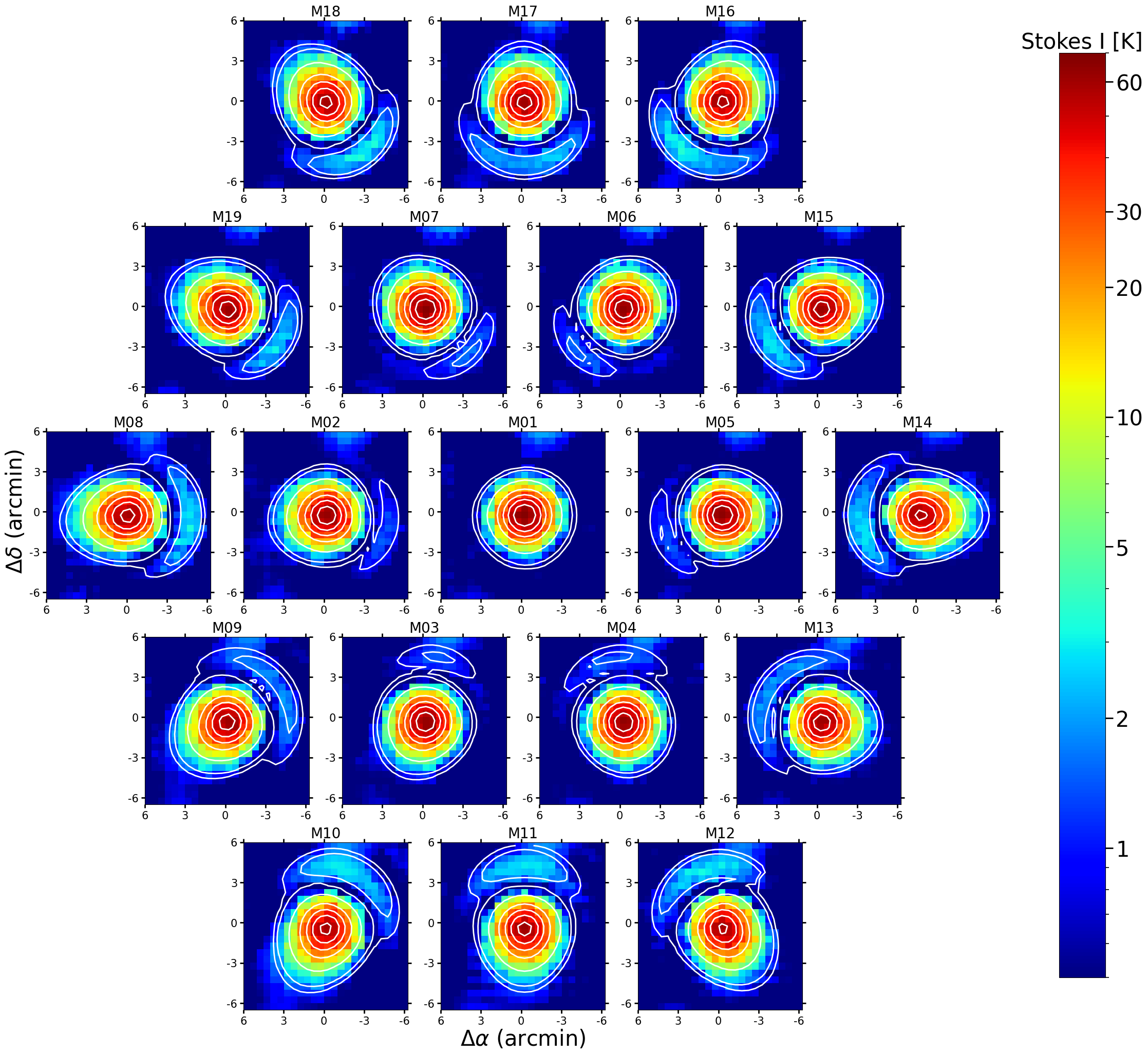}
\caption{Stokes $I$ mapping of J0854+2006 for the 19 beams near 1420 MHz. To compare with the simulated beam structures (as shown in Figure \ref{fig7}), the mapping data has been mirror-symmetrized. The feature near the source of each map is attributed to a nearby secondary source, J0854+1959. White solid lines indicate contour levels at 1\%, 2\%, 10\%, 30\%, 50\%, 70\%, and 90\% of the peak flux.}
\label{Fig6}
\end{figure*}

Analogous to Figure \ref{Fig2}, Figure \ref{Fig4} illustrates the fitting outcome for beam M08 of Stokes $I$. The discrepancy is negligible, with no discernible residual structure, indicating a satisfactory fitting to the asymmetrical beam shape and sidelobe. Furthermore, Figure \ref{Fig5} presents the cross sections along the major and minor axes for beam M08. The major axis showcases an asymmetrical distribution and response from the sidelobe, effectively fitted by the aforementioned method. These findings corroborate that the log-normal distribution combined with the Gaussian distribution is suitable for the off-center beams, and the response from the sidelobes exhibits a Gaussian profile. This asymmetry can affect the accuracy of data and needs to be accounted for in analyses. 

\begin{figure*}
\centering
\includegraphics[width=\textwidth, angle=0]{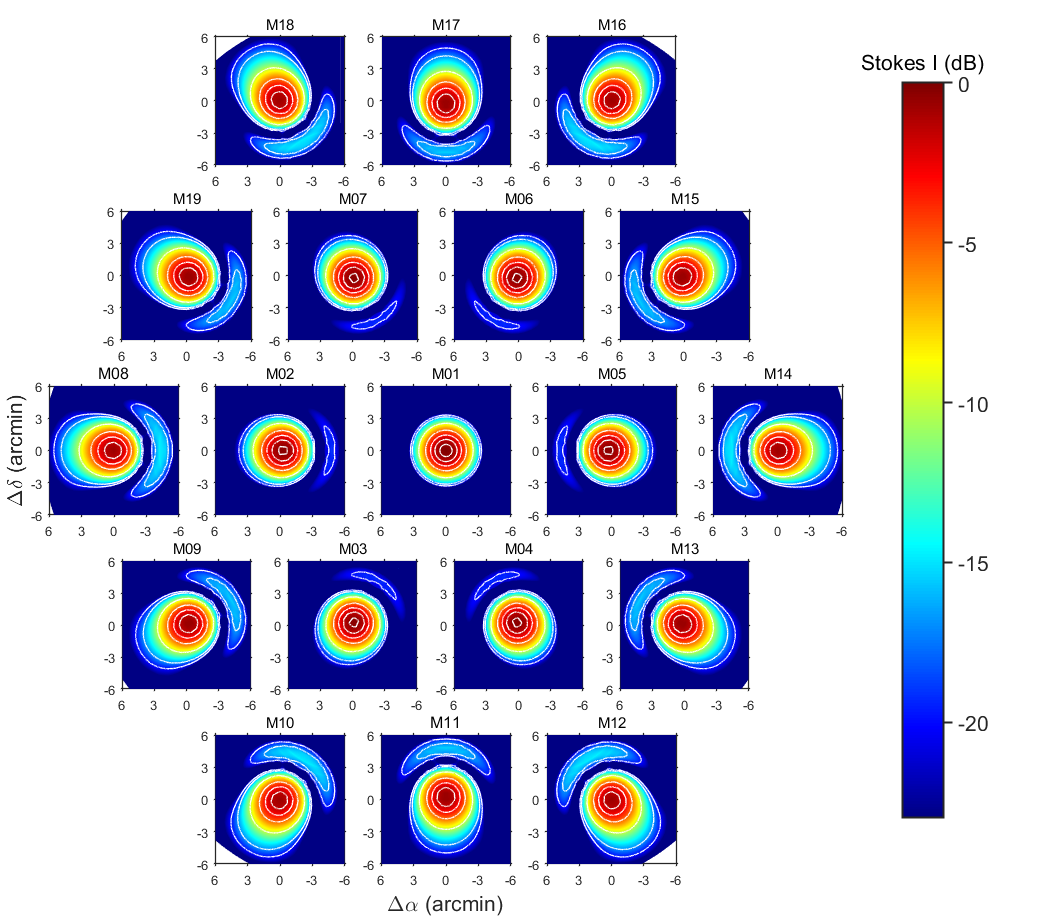}
\caption{Simulated Stokes $I$ patterns of all 19 individual beams.}
\label{fig7}
\end{figure*}

\begin{figure*}
\centering
\begin{minipage}{0.49\textwidth}
    \centering
    \includegraphics[width=\textwidth, angle=0]{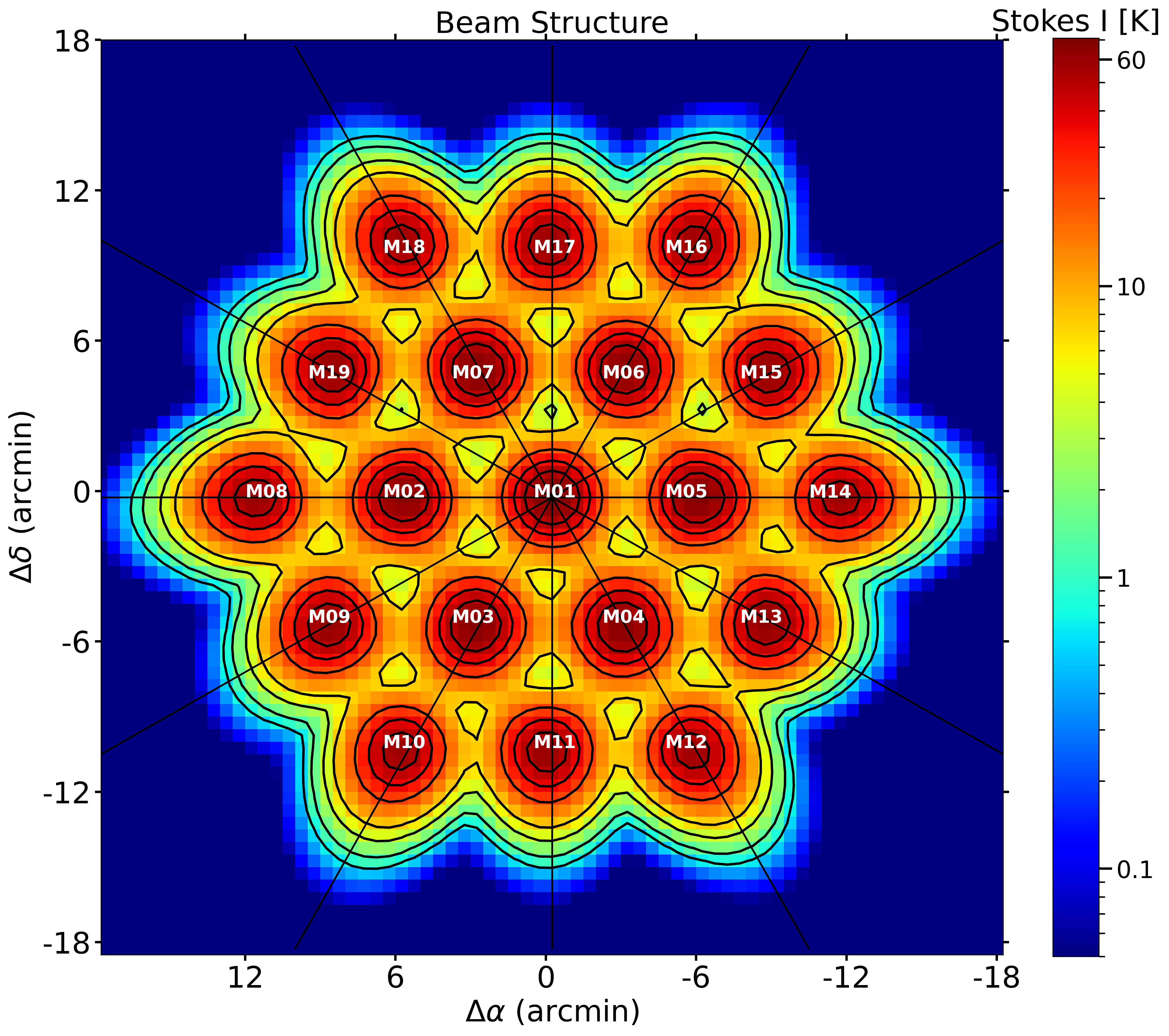}
\end{minipage}%
\hfill
\begin{minipage}{0.49\textwidth}
    \centering
    \includegraphics[trim=.7cm 0cm 1.2cm 0cm,clip,width=\textwidth, angle=0]{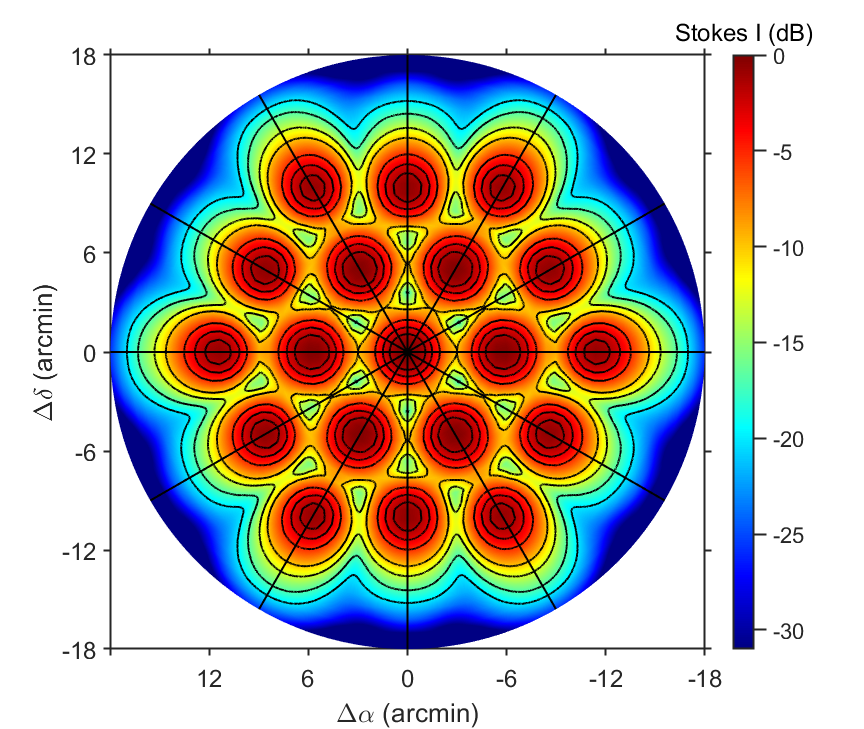}
\end{minipage}
\caption{\textbf{Left}: Fitted combined response of the 19 beams based on observation of J0854+2006 in Stokes $I$. The black solid lines are spaced 30 degrees apart. Contours (black lines) depict intensities at levels of 1\%, 2\%, 5\%, 10\%, 30\%, 50\%, and 70\% of the peak flux, respectively. The $x$ and $y$ axes represent the RA and Dec offsets relative to the center of beam M01 (see Table \ref{tab2}). \textbf{Right}: Simulated total intensity beam pattern of the 19 beams combined.}
\label{fig8}
\end{figure*}

\begin{figure*}
\centering
\begin{minipage}{0.8\textwidth}
    \centering
    \includegraphics[width=\textwidth, angle=0]{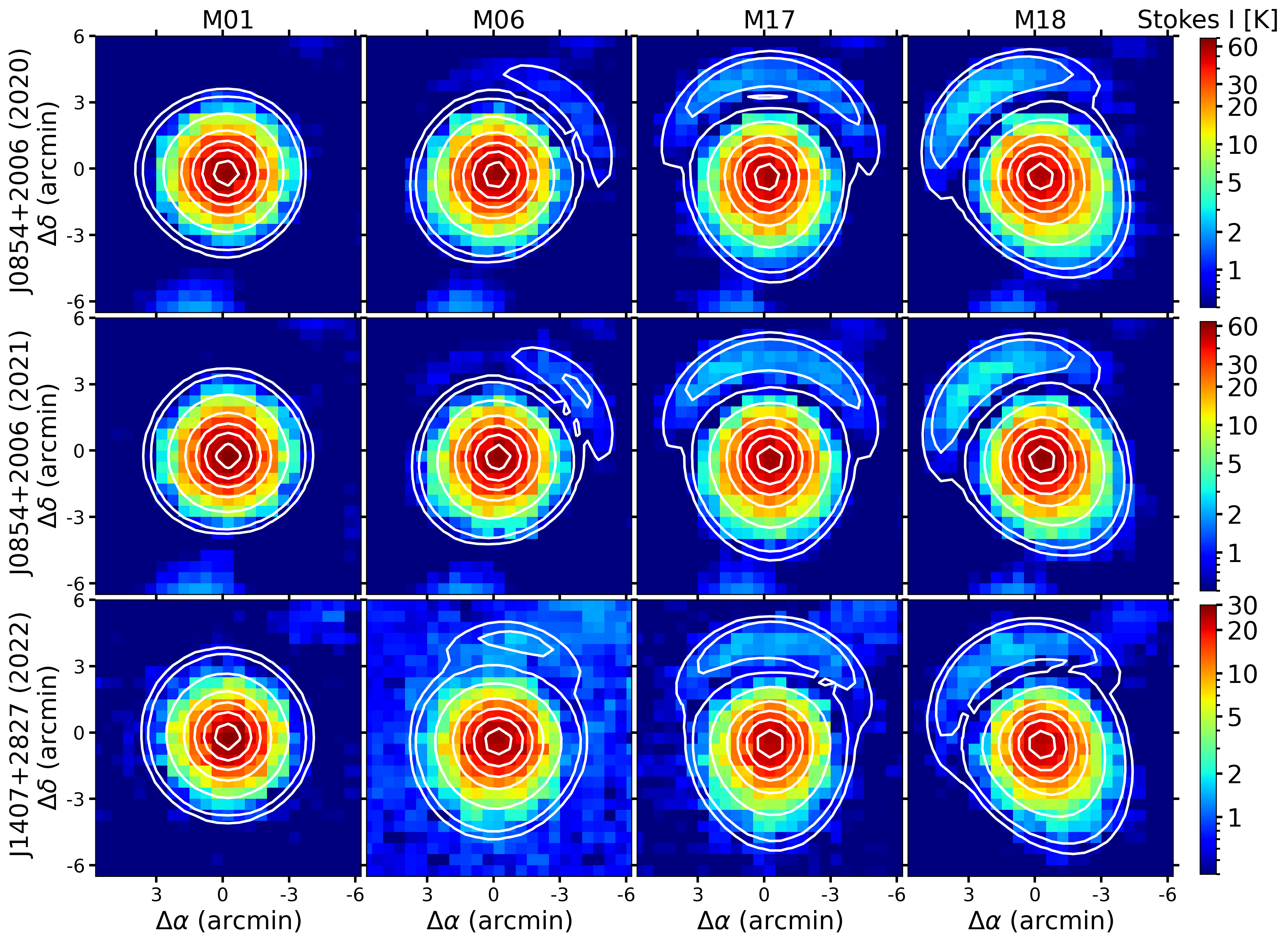}
\end{minipage}%
\vfill
\begin{minipage}{0.8\textwidth}
    \centering
    \includegraphics[trim=0.8cm 0cm 6cm 0cm,clip,width=\textwidth, angle=0]{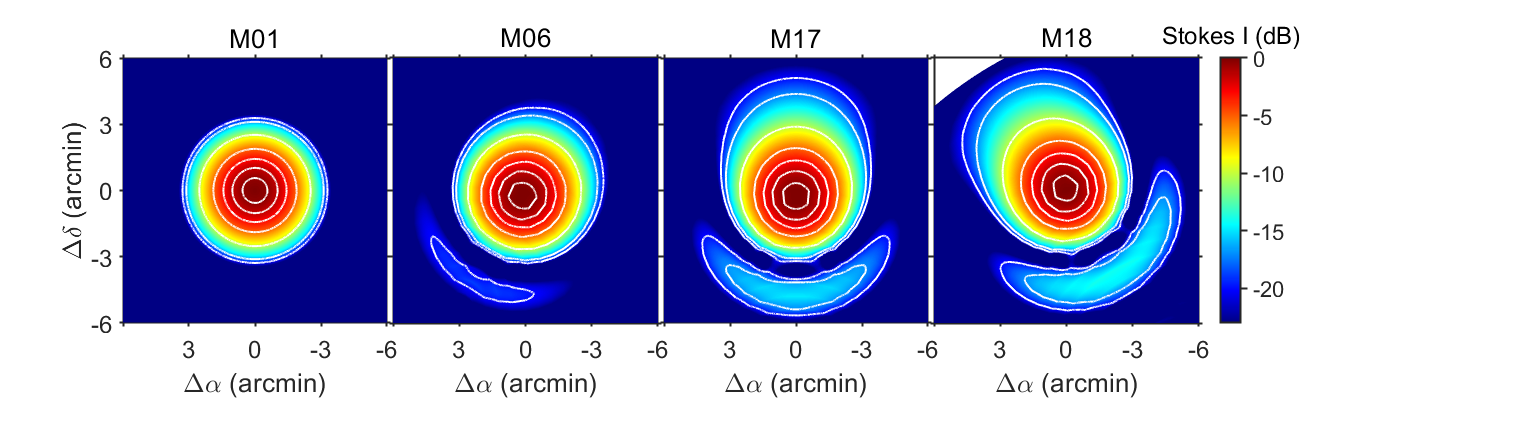}
\end{minipage}
\caption{\textbf{Top}: Stokes $I$ mapping of J0854+2006 (2020), J0854+2006 (2021), and J1407+2827 (2022) for beams M01, M06, M017, and M18. The feature near the bottom of J0854+2006 is attributed to a nearby secondary source, J0854+1959. Contours (white lines) represent the fitting results, showcasing intensities at levels of 1\%, 2\%, 10\%, 30\%, 50\%, 70\%, and 90\%, respectively. \textbf{Bottom}: Simulated total intensity beam patterns of M01, M06, M17, and M18. Contours are of same intensity levels.}
\label{fig9}
\end{figure*}

\begin{figure*}
\centering
\includegraphics[width=\textwidth, angle=0]{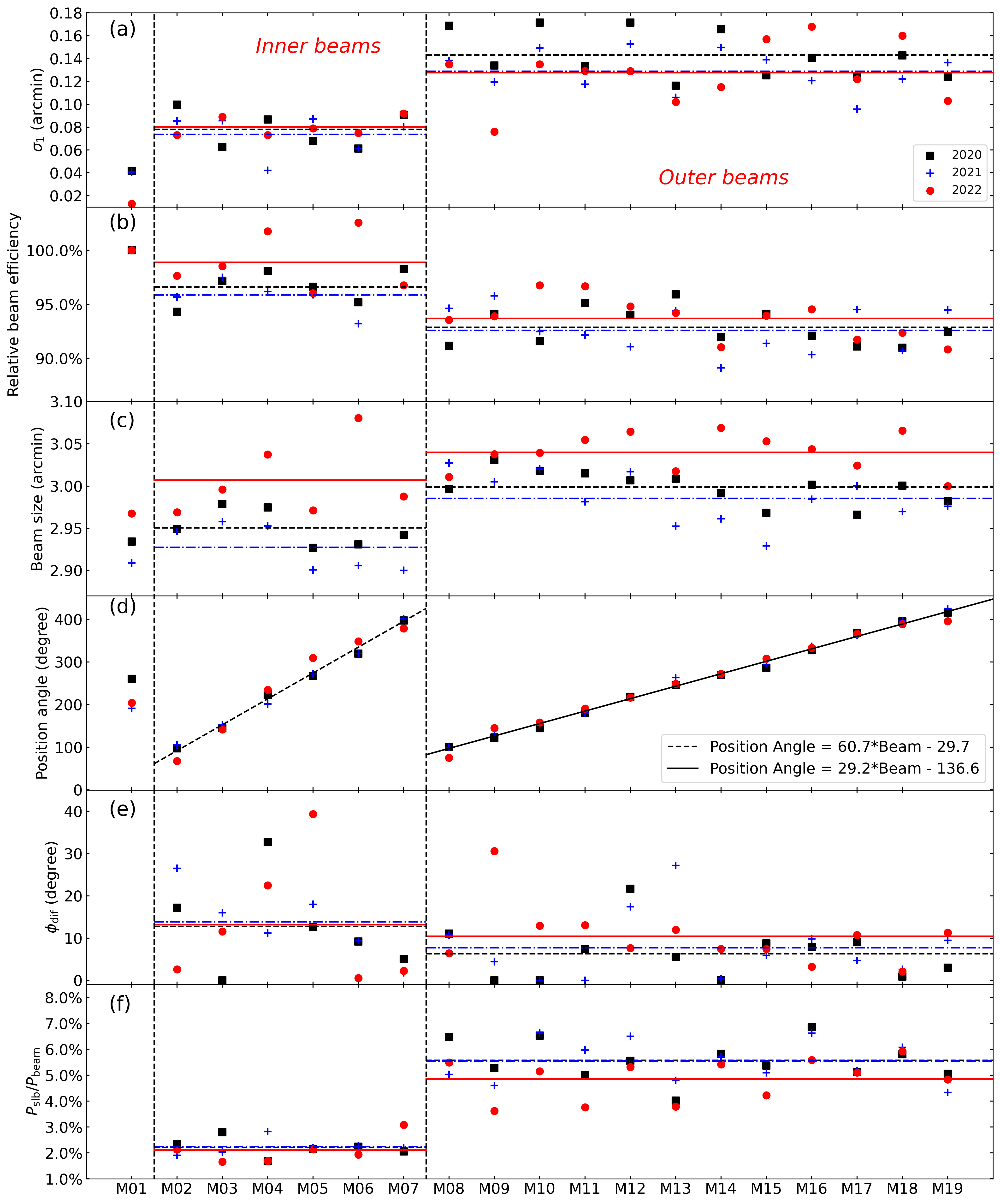}
\caption{Comparative analysis of Stokes $I$ parameters across various observations in inner and outer beams of the FAST Telescope. Parameters include $\sigma_1$, the relative beam efficiency, beam size, position angle, $\phi_{\rm dif}$, and $P_{slb}/P_{beam}$. Inner beams: M02--M07; outer beams: M08--M19. Data points from 2022, 2021, and 2020 are denoted by red circles, blue plus signs, and black squares, respectively. The red solid, black dashed, and blue dash-dotted lines indicate the average values of the corresponding parameters across all observations, respectively. In panel (d), the black solid and black dashed lines represent the linear fitting results of the position angle distribution for the outer and inner beams, respectively.}
\label{fig10}
\end{figure*}

\subsection{Stokes $I$ Parameterization}
The combined log-normal and Gaussian distribution approach was applied to analyze 19 beams over three years (for the central beam, both the Gaussian-only and log-normal-and-Gaussian methods are applied). The analysis included determining fitting parameters and computing characteristics such as the position of the beam center, beam size, the relative beam efficiency, and the ratio of the sidelobe intensity contribution to the main lobe intensity ($P_{\rm slb}/P_{\rm beam}$). Regarding the elongated asymmetrical axis of the main beam as the major axis, we calculate the FWHM for the major and minor axes using the following formulas:
\begin{equation}
\mathrm{FWHM_{major}} = e^{\mu_1 + \sigma_1 \sqrt{2 \ln 2}} - e^{\mu_1 - \sigma_1 \sqrt{2 \ln 2}} \, ,
\label{eq:FWHM_major}
\end{equation}
\begin{equation}
\mathrm{FWHM_{minor}} = 2 \sqrt{2 \ln 2} \cdot \sigma \, .
\label{eq:FWHM_minor}
\end{equation}
Subsequently, the beam size is derived as the geometric mean of $\mathrm{FWHM_{major}}$ and $\mathrm{FWHM_{minor}}$, given by:
\begin{equation}
\text{Beam size} = \sqrt{\mathrm{FWHM_{major}} \cdot \mathrm{FWHM_{minor}}}.
\label{eq:beam_size}
\end{equation} 
The relative beam efficiency of the off-center beams with respect to the central beam is determined by comparing the ratio of the total intensity of each beam to that of the central beam. The ratio $P_{\rm slb}/P_{\rm beam}$ is obtained by comparing the total intensity of the sidelobes to the total intensity of the main lobe. All the fitted parameters are listed in Table \ref{tab1}, while other parameters calculated based on the fitted values (e.g., beam center, beam size) are shown in Table \ref{tab2}.

Figure \ref{Fig6} displays Stokes $I$ maps for each individual beam based on 2021 observations of J0854+2006, while Figure \ref{fig7} presents the corresponding simulated results. The coma and sidelobes are clearly displayed and perfectly fitted, with the outer beams showing more prominent coma and sidelobes compared to the inner beams. Additionally, the position angles of the beams follow a regular pattern, with the orientation of the major axis of each main lobe and the orientation of each sidelobe being largely consistent. Furthermore, the coma all extend outward from the central beam, indicating a systematic variation in the position angle of the main lobe. Overall, the observational results closely align with the simulations, particularly regarding the structures of the coma and sidelobes. Figure \ref{fig8} presents the fitted results (left) and the simulations (right) for the combined response of the 19 beams, further demonstrating a high degree of consistency. The outer sidelobe structures reported by \citet{jiang2020} are absent. This is because the raster scan mapping is a mirrored image of the beam pattern.

In general, the 19 beams of the FAST telescope can be categorized into four groups based on the exact angular distance of each beam's center from that of the central beam M01, denoted as $d_{\rm M01}$. For simplicity, we classify the beams into three groups: (1) the central beam M01; (2) inner beams with relatively small $d_{\rm M01}$ ($\leq$ 7$^\prime$), comprising M02, M03, M04, M05, M06, and M07; (3) outer beams with larger $d_{\rm M01}$  ($\geq$ 7$^\prime$), comprising M08, M09, M10, M11, M12, M13, M14, M15, M16, M17, M18, and M19. For beams M01, M06, M17, and M18, each characterized by distinct $d_{\rm M01}$ values, we present the distribution maps of total intensity with fitting results using a logarithmic-scale plot to explore the relationships between asymmetrical structures and $d_{\rm M01}$, as depicted in Figure \ref{fig9}. Three years of observations yield consistent results: For beam M01, no asymmetrical structures were detected. In contrast, the asymmetrical features of the sidelobes and the coma begin to appear in M06 and become more prominent in M17 and M18, indicating that the asymmetrical structures become more pronounced as $d_{\rm M01}$ increases. However, the position angle of the M06 sidelobe in 2022 differs significantly from the results in 2020 and 2021. This variation may be due to the different zenith angles of the two sources or a time variability of the feed from 2021 to 2022. The coma and sidelobes are oriented in opposite directions. Since these features are intrinsic to the laterally displaced horn, they are expected to remain stable over time. These results are in agreement with the simulation, indicating that the farther from the focus, the more pronounced the coma and sidelobes become.

Figure \ref{fig10} provides a comparative analysis of Stokes $I$ parameters across various observations. Figure \ref{fig10}(a) illustrates the distribution of $\sigma_1$, where larger values of $\sigma_1$ indicate greater deviation from the Gaussian distribution, corresponding to more pronounced coma and higher levels of asymmetry. Over the three-year epochs, outer beams consistently exhibit higher levels of asymmetry compared to inner beams. This is consistent with the results of \citet{jiang2020}, who proposed that notable features become more significant with increasing $d_{\rm M01}$ for the outer beams. Figure \ref{fig10}(b) delineates the variation in the relative beam efficiency across all observations. The majority of inner beams (M01 to M07) exhibit high efficiencies, averaging $\geq 95\%$, while outer beams demonstrate relatively lower efficiencies, averaging between $95\%$ to $90\%$. In Figure \ref{fig10}(c), the beam size of all beams is depicted. The beam size estimated based on observations of J1407+2827 consistently exceeds that based on J0854+2006. Inner beams consistently demonstrate relatively smaller beam sizes compared to outer beams, with a difference not exceeding 0.1 arcmin. Additionally, noticeable disparities in beam size are observed when observing different sources, while temporal variations in beam size are negligible when observing the same source. Moreover, the average difference between the major and minor beam widths is 0.1/0.2 arcmin for inner/outer beams, contrasting with \citet{Sun2021}, who proposed that the difference is very small. These findings underscore the superior performance of inner beams in terms of both beam efficiency and size relative to outer beams. Interestingly, in the case of the same source J0854+2006, when comparing data between 2020 and 2021, it is noted that the asymmetry level and beam size of the outer beams are systematically smaller in the 2021 results, which indicates that the performance of the outer beams in 2021 may have surpassed that of 2020.

Disparities in position angle across different observations are illustrated in Figure \ref{fig10}(d). The position angle distribution shows a high degree of consistency over three years of observations. For the inner beams, the position angles follow a linear distribution with a slope of approximately 60.7 degrees, while the outer beams exhibit a linear distribution with a slope of approximately 29.2 degrees. This result is a direct consequence of the hexagonal layout of the feedhorns, as shown by the black lines in Figure \ref{fig8}. This layout is a common design choice for receiver arrays, such as the 7-pixel L-band array used at Arecibo \citep{Giovanelli2005}. To better understand the relationship between sidelobes and beams, we calculate the angle difference ($\phi_{\rm dif} = \phi - \phi_c$), representing the angle between the major axis of the beam and the symmetry axis of the sidelobe. Figure \ref{fig10}(e) shows the distribution of $\phi_{\rm dif}$, with most beams exhibiting $\phi_{\rm dif} \leq 20^{\circ}$, indicating a strong correlation between the location of sidelobes and the position angle of the beam. This correlation is also evident in Figure \ref{fig8}, where black lines intersect the center of sidelobes. These results indicate that the coma and sidelobes largely determine the position angle of each beam, making the position angle stable and solely dependent on the design of the lateral offset of the horns in FAST, without varying over time.

Of particular importance is the quantification of the percentage of flux contributed by sidelobes. By successfully modeling sidelobes of each beam, we can now reliably quantify this parameter, denoted by $P_{\rm slb}/P_{\rm beam}$, representing the ratio of the total intensity of sidelobes to the total intensity of the main lobe, as depicted in Figure \ref{fig10}(f). sidelobes of inner beams contribute nearly 2\% of the total flux of the main lobe, while this contribution increases to 5\% for outer beams, peaking at 6.8\% for M16. These results indicate that sidelobes may significantly impact related studies, such as the determination of the position of Fast Radio Bursts (FRBs) and the H\uppercase\expandafter{\romannumeral1} intensity mapping survey. Further research is warranted to quantify these impacts stemming from sidelobes.

In Figure \ref{fig10}, the black dashed, blue dash-dotted, and red solid lines indicate the average values of the corresponding parameters in 2020, 2021, and 2022, respectively. In most panels, the black dashed lines and blue dash-dotted lines have similar values, but the deviations of the red solid lines from the black dashed lines and blue dash-dotted lines are statistically significant compared to the scattering of the data, especially the inner and outer beams of Figure \ref{fig10}(c). These deviations of the 2022 results from the 2020 and 2021 results may indicate a dependence of beam structure on source declination or a time variability of the receiving system.

\newpage 
\begin{sidewaystable*} 
  \vspace*{7cm}
  \centering
  \caption{Fitting parameters of the 19 beams in Stokes $I$ obtained by applying combined log-normal and Gaussian fitting to the observed total intensity maps towards J0854+2006. The Gaussian fitting result for the central beam is also listed. Uncertainties are the square roots of the diagonal elements of the covariance matrix. Units for \( A \) are in kelvin, and for \( \mu_1 \), \( \mu_2 \), \( \sigma_1 \), \( \sigma_2 \), \( a \), \( r_c \), \( \phi_c \) are in arcmin.}
  \label{tab1}
  \renewcommand\tabcolsep{4.0pt} %
  \begin{tabular}{cccccccccccccc}
  \hline
     Beam & \( A \) & \( \mu_1 \) & \( \mu_2 \) & \( \sigma_1 \) & \( \sigma_2 \) & \( \phi \) & \( a \) & \( r_c \) & \( \phi_c \) & \( \sigma_3 \) & \( \sigma_4 \) & \( A_1 \) \\
    & (K) & (arcmin) & (arcmin) & (arcmin) & (arcmin) & (deg) & (arcmin) & (arcmin) & (deg) & (arcmin) & (arcmin) & (K) \\
  \hline
     M01 (Gauss)      &	655$\pm5$	&	0.05$\pm0.01$	&	0.01$\pm0.01$	&	1.31$\pm0.01$	&	1.29$\pm0.01$	&	141$\pm27$ 	&	-	&	-	&	-	&	-	&	-	&	-	\\											
   \hline																																			
    M01	&	22$\pm$4	&	3.4$\pm$0.2	&	0.00$\pm$0.01 	&	0.04$\pm$0.01 	&	1.25$\pm$0.01 	&	191$\pm$5.3 	&	30.0$\pm$5.6 	&	-			&	-			&	-			&	-			&	-			\\
    M02	&	43$\pm$4 	&	2.7$\pm$0.1	&	0.00$\pm$0.01 	&	0.09$\pm$0.01 	&	1.24$\pm$0.01 	&	105$\pm$3	&	14.7$\pm$1.4 	&	4.7$\pm$0.2	&	101$\pm$16 	&	0.5$\pm$0.2	&	0.68$\pm$0.30 	&	2.6$\pm$1.1	\\										
    M03	&	43$\pm$4	&	2.7$\pm$0.1	&	0.00$\pm$0.01 	&	0.09$\pm$0.01 	&	1.23$\pm$0.01 	&	153$\pm$3	&	14.8$\pm$1.5 	&	4.9$\pm$0.2	&	169$\pm$10	&	0.6$\pm$0.3	&	0.49$\pm$0.18 	&	2.7$\pm$1.1	\\										
    M04	&	21$\pm$4	&	3.4$\pm$0.2	&	0.00$\pm$0.02 	&	0.04$\pm$0.01 	&	1.24$\pm$0.01 	&	202$\pm$5	&	30.0$\pm$6.0 	&	4.8$\pm$0.2	&	190$\pm$13	&	0.6$\pm$0.2	&	0.70$\pm$0.24 	&	3.9$\pm$1.3	\\										
    M05	&	44$\pm$4	&	2.7$\pm$0.1	&	0.00$\pm$0.01 	&	0.09$\pm$0.01 	&	1.21$\pm$0.01 	&	272$\pm$3	&	14.4$\pm$1.3 	&	4.7$\pm$0.1 	&	290$\pm$9	&	0.6$\pm$0.1 	&	0.65$\pm$0.17 	&	3.0$\pm$0.8	\\										
    M06	&	31$\pm$4	&	3.0$\pm$0.1	&	0.00$\pm$0.01 	&	0.06$\pm$0.01 	&	1.23$\pm$0.01 	&	320$\pm$5	&	20.2$\pm$2.8 	&	4.7$\pm$0.1	&	310$\pm$7	&	0.6$\pm$0.1	&	0.58$\pm$0.12 	&	3.0$\pm$0.6	\\										
    M07	&	42$\pm$4	&	2.7$\pm$0.1	&	0.00$\pm$0.01 	&	0.08$\pm$0.01 	&	1.23$\pm$0.01 	&	401$\pm$3	&	15.3$\pm$1.5 	&	4.7$\pm$0.1	&	402$\pm$6	&	0.6$\pm$0.1	&	0.58$\pm$0.11 	&	3.1$\pm$0.6	\\										
    M08	&	64$\pm$4	&	2.3$\pm$0.1	&	0.00$\pm$0.01 	&	0.14$\pm$0.01 	&	1.22$\pm$0.01 	&	99$\pm$2	&	9.6$\pm$0.6 	&	4.6$\pm$0.1	&	92$\pm$5	&	0.6$\pm$0.1 	&	0.73$\pm$0.08 	&	7.1$\pm$0.8	\\										
    M09	&	58$\pm$4	&	2.4$\pm$0.1	&	0.00$\pm$0.01 	&	0.12$\pm$0.01 	&	1.24$\pm$0.01 	&	132$\pm$2	&	10.8$\pm$0.8 	&	4.7$\pm$0.1	&	128$\pm$7	&	0.6$\pm$0.1	&	0.94$\pm$0.15 	&	7.0$\pm$1.0	\\										
    M10	&	68$\pm$5	&	2.2$\pm$0.1	&	0.00$\pm$0.01 	&	0.15$\pm$0.01 	&	1.23$\pm$0.01 	&	153$\pm$2	&	8.7$\pm$0.6 	&	4.6$\pm$0.1	&	153$\pm$4	&	0.7$\pm$0.1	&	0.72$\pm$0.08 	&	9.2$\pm$1.0	\\										
    M11	&	55$\pm$4	&	2.4$\pm$0.1	&	0.00$\pm$0.01 	&	0.12$\pm$0.01 	&	1.24$\pm$0.01 	&	180$\pm$2	&	10.8$\pm$0.8 	&	4.7$\pm$0.1	&	180$\pm$4	&	0.8$\pm$0.1	&	0.58$\pm$0.06 	&	7.9$\pm$0.8	\\										
    M12	&	69$\pm$4	&	2.2$\pm$0.1	&	0.03$\pm$0.01 	&	0.15$\pm$0.01 	&	1.23$\pm$0.01 	&	215$\pm$2	&	8.5$\pm$0.6 	&	4.5$\pm$0.1	&	197$\pm$4	&	0.6$\pm$0.1	&	0.78$\pm$0.08 	&	9.7$\pm$0.9	\\										
    M13	&	52$\pm$4	&	2.5$\pm$0.1	&	0.15$\pm$0.01 	&	0.11$\pm$0.01 	&	1.25$\pm$0.01 	&	263$\pm$2	&	11.7$\pm$0.9 	&	4.6$\pm$0.1 	&	236$\pm$6	&	0.6$\pm$0.1	&	1.00$\pm$0.14 	&	7.5$\pm$0.9	\\										
    M14	&	66$\pm$4	&	2.2$\pm$0.1	&	0.01$\pm$0.01 	&	0.15$\pm$0.01 	&	1.19$\pm$0.01 	&	270$\pm$1	&	8.7$\pm$0.5 	&	4.5$\pm$0.1	&	270$\pm$4	&	0.6$\pm$0.1	&	0.73$\pm$0.08 	&	7.8$\pm$0.8	\\										
    M15	&	66$\pm$4	&	2.2$\pm$0.1	&	0.00$\pm$0.01 	&	0.14$\pm$0.01 	&	1.22$\pm$0.01 	&	294$\pm$2 	&	9.0$\pm$0.6 	&	4.6$\pm$0.1	&	300$\pm$3	&	0.6$\pm$0.1	&	0.60$\pm$0.05 	&	6.8$\pm$0.6	\\										
    M16	&	56$\pm$4	&	2.4$\pm$0.1	&	0.00$\pm$0.01 	&	0.12$\pm$0.01 	&	1.24$\pm$0.01 	&	336$\pm$2	&	10.5$\pm$0.8 	&	4.6$\pm$0.1 	&	326$\pm$4	&	0.6$\pm$0.1	&	0.78$\pm$0.08 	&	9.2$\pm$0.9	\\										
    M17	&	46$\pm$4	&	2.6$\pm$0.1	&	0.00$\pm$0.01 	&	0.10$\pm$0.01 	&	1.26$\pm$0.01 	&	363$\pm$3	&	13.3$\pm$1.3 	&	4.6$\pm$0.1	&	358$\pm$6	&	0.6$\pm$0.1	&	0.89$\pm$0.12 	&	7.6$\pm$0.9	\\										
    M18	&	56$\pm$4	&	2.4$\pm$0.1	&	0.03$\pm$0.01 	&	0.12$\pm$0.01 	&	1.22$\pm$0.01 	&	397$\pm$2	&	10.5$\pm$0.8 	&	4.5$\pm$0.1	&	394$\pm$3	&	0.6$\pm$0.1	&	0.65$\pm$0.05 	&	8.2$\pm$0.6	\\										
    M19	&	66$\pm$4	&	2.2$\pm$0.1	&	0.01$\pm$0.01 	&	0.14$\pm$0.01 	&	1.23$\pm$0.01 	&	425$\pm$2	&	9.4$\pm$0.6 	&	4.6$\pm$0.1	&	416$\pm$4	&	0.6$\pm$0.1	&	0.58$\pm$0.07 	&	6.0$\pm$0.7	\\																			
  \hline
  \end{tabular}
\end{sidewaystable*}

\begin{table}
  \centering
  \scriptsize
  \caption{Beam center, beam size, relative beam efficiency, and the ratio of sidelobe contribution to main lobe contribution, $P_{\rm slb}/P_{\rm beam}$, are calculated based on Table \ref{tab1}.}
  \label{tab2}
  \renewcommand\tabcolsep{4.0pt} %
  \begin{tabular}{crrccc}
  \hline
   Beam	&$X_c$	&	$Y_c$	&	Beam Size	&	Efficiency	&	$f_{\rm slb}$	\\
    & $(^\prime)$& $(^\prime)$& $(^\prime)$	& & \\
  \hline
 M01	&	0.08 	&	0.00 	&	2.91 	&	100\%	&	-	\\
    M02	&	5.81 	&	0.07 	&	2.95 	&	96\%	&	1.91\%	\\
    M03	&	2.85 	&	-4.92 	&	2.96 	&	97\%	&	2.04\%	\\
    M04	&	-2.88 	&	-4.98 	&	2.95 	&	96\%	&	2.83\%	\\
    M05	&	-5.62 	&	0.03 	&	2.90 	&	96\%	&	2.21\%	\\
    M06	&	-2.73 	&	4.97 	&	2.91 	&	93\%	&	2.26\%	\\
    M07	&	3.04 	&	4.98 	&	2.90 	&	97\%	&	2.21\%	\\
    M08	&	11.70 	&	0.07 	&	3.03 	&	95\%	&	5.03\%	\\
    M09	&	8.67 	&	-4.78 	&	3.01 	&	96\%	&	4.61\%	\\
    M10	&	5.84 	&	-9.86 	&	3.02 	&	92\%	&	6.63\%	\\
    M11	&	0.04 	&	-9.90 	&	2.98 	&	92\%	&	5.98\%	\\
    M12	&	-5.49 	&	-9.96 	&	3.02 	&	91\%	&	6.50\%	\\
    M13	&	-8.41 	&	-4.99 	&	2.95 	&	94\%	&	4.80\%	\\
    M14	&	-11.30 	&	-0.03 	&	2.96 	&	89\%	&	5.73\%	\\
    M15	&	-8.41 	&	4.95 	&	2.93 	&	91\%	&	5.10\%	\\
    M16	&	-5.54 	&	9.94 	&	2.98 	&	90\%	&	6.62\%	\\
    M17	&	0.22 	&	9.93 	&	3.00 	&	95\%	&	5.16\%	\\
    M18	&	6.01 	&	10.01 	&	2.97 	&	91\%	&	6.08\%	\\
    M19	&	8.81 	&	4.98 	&	2.98 	&	94\%	&	4.34\%	\\
  \hline

  \end{tabular}
\end{table}

\begin{figure*}
\centering
\begin{minipage}{\textwidth}
    \centering
    \includegraphics[width=0.8\textwidth, angle=0]{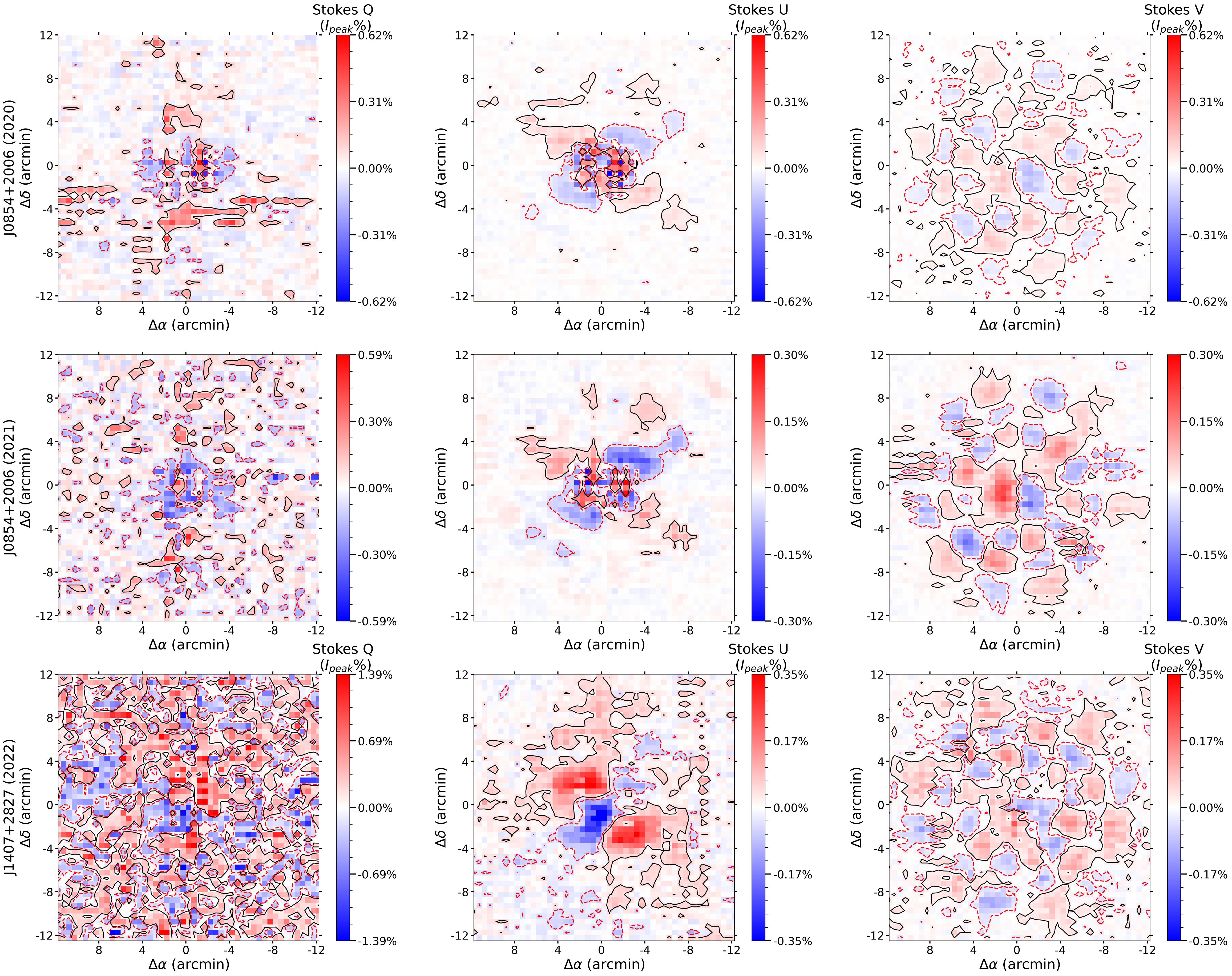}
\end{minipage}%
\vfill
\begin{minipage}{\textwidth}
    \centering
    \includegraphics[trim = 0cm 0cm 0cm 0cm,clip,width=\textwidth, angle=0]{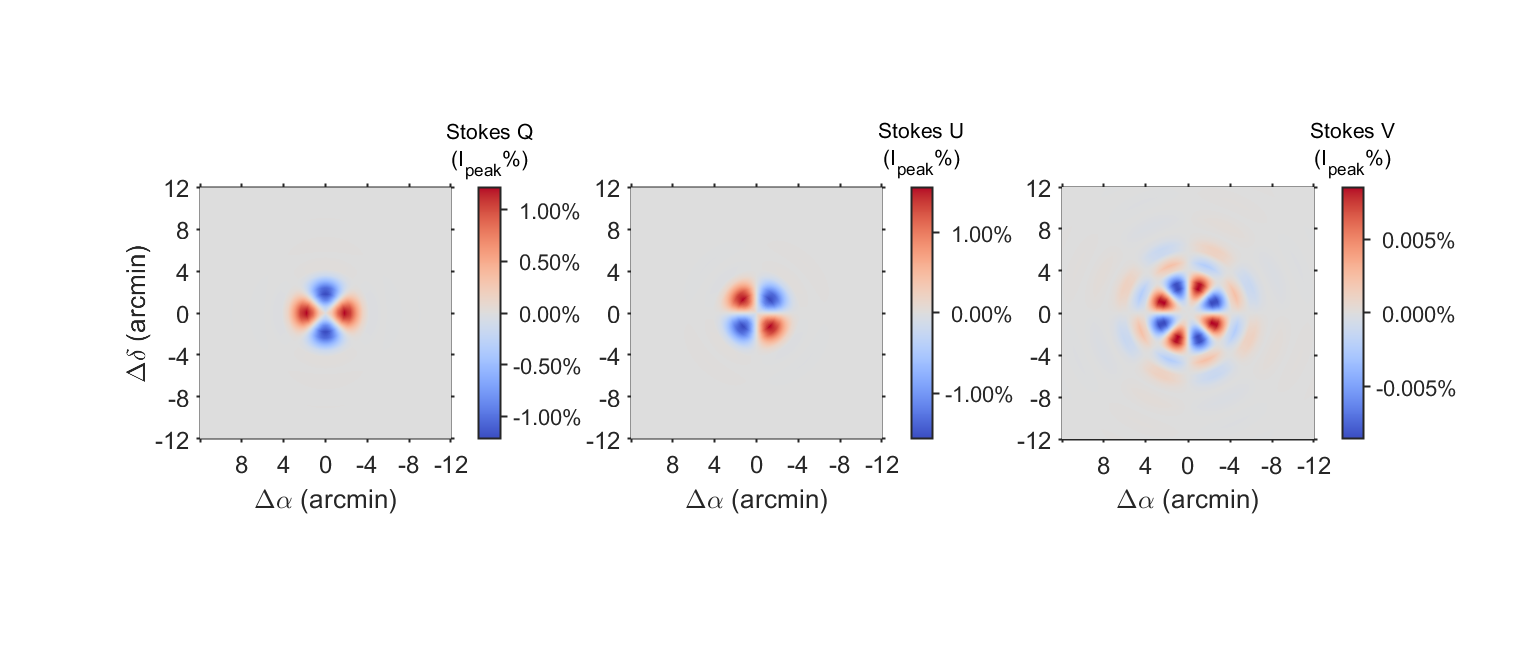}
\end{minipage}
\caption{\textbf{Top:} Mapping of Stokes $Q$, $U$, and $V$ for the central beam across different observations. Contours (black solid lines) represent the fractional polarization $Q/I_{peak}$, $U/I_{peak}$, $V/I_{peak}$ at 0.1\%, 0.04\%, 0.02\%, respectively, while contours (red dashed lines) represent the fractional polarization $Q/I_{peak}$, $U/I_{peak}$, $V/I_{peak}$ at $-0.1\%$, $-0.02\%$, $-0.01\%$, respectively. The colorbar is marked with the fractional polarization relative to the peak of Stokes $I$ from the central beam, denoted as $I_{\rm peak}\%$ (64.44 K, 67.50 K, and 28.85 K for data observed in 2020, 2021, and 2022, respectively). For the data observed in the same year, the colorbar range for Stokes $U$ and $V$ is kept consistent for comparison purposes. \textbf{Bottom}: Simulated Stokes $Q$, $U$, and $V$ patterns of the central beam.}
\label{fig11}
\end{figure*}

\begin{figure*}
\centering
\includegraphics[width=\textwidth, angle=0]{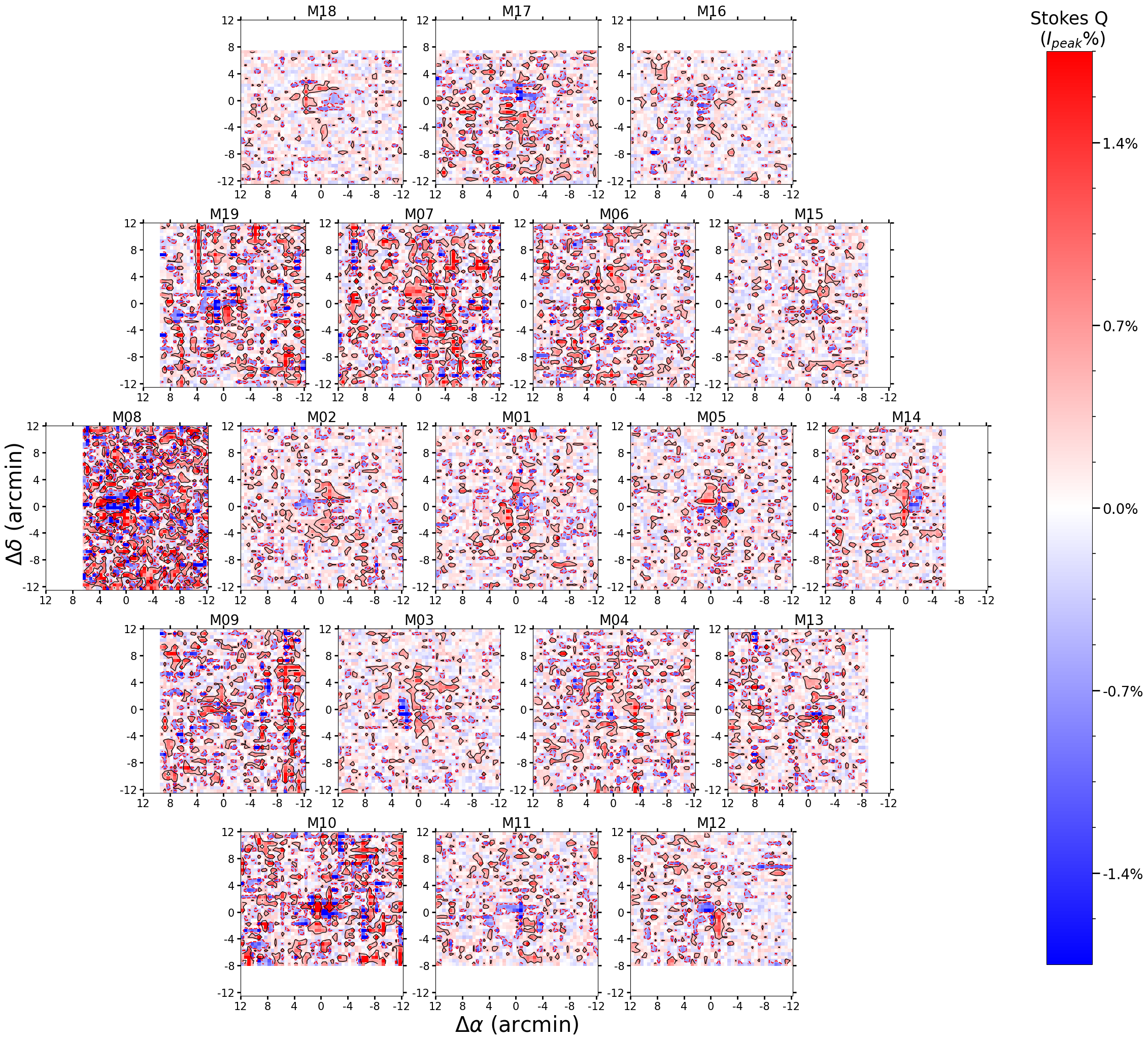}
\caption{Stokes $Q$ mapping of J1407+2827 for the 19 beams near 1420 MHz. To compare with the simulated beam patterns, the mapping data have been mirror-symmetrized. Black solid contours represent $Q/I_{peak}$ at 0.4\%, while red dashed contours represent $Q/I_{peak}$ at $-0.4\%$.} The blank areas in the outer beams are outside the scanning region.
\label{fig12}
\end{figure*}

\begin{figure*}
\centering
\includegraphics[width=\textwidth, angle=0]{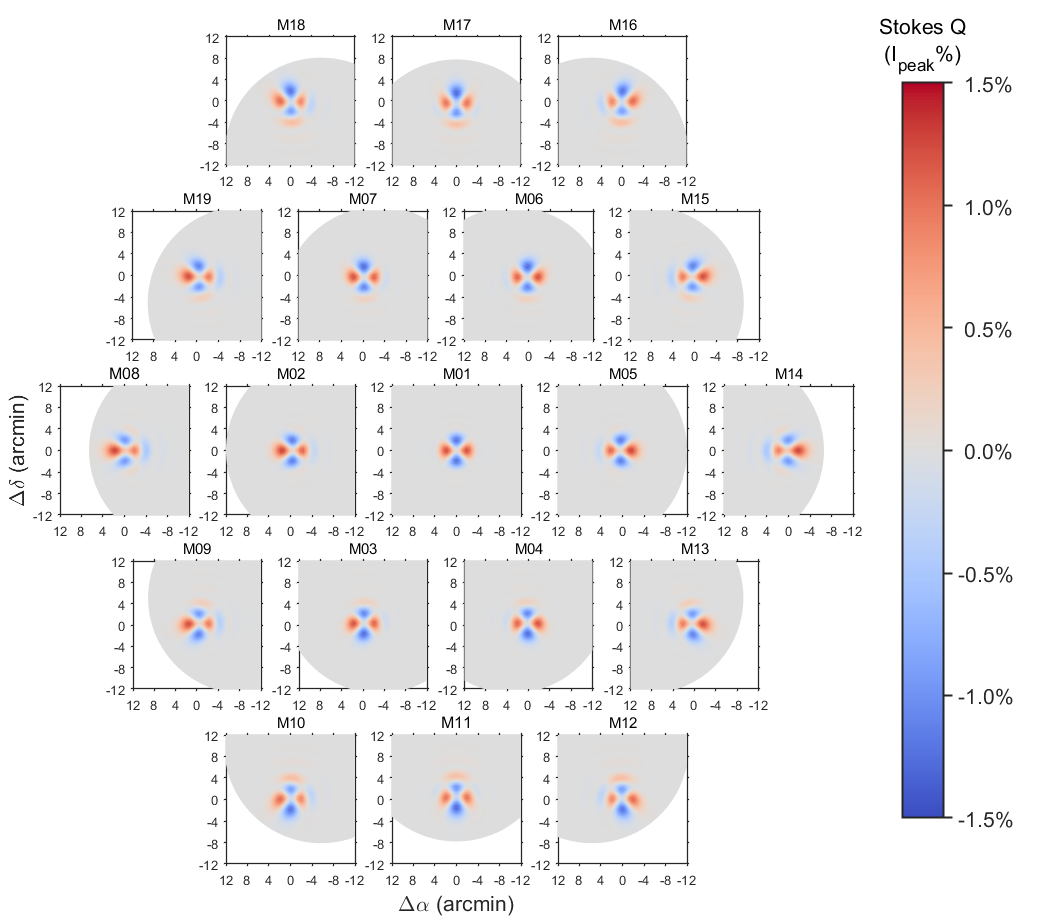}
\caption{Simulated Stokes $Q$ patterns of all 19 individual beams.}
\label{fig13}
\end{figure*}

\begin{figure*}
\centering
\includegraphics[width=\textwidth, angle=0]{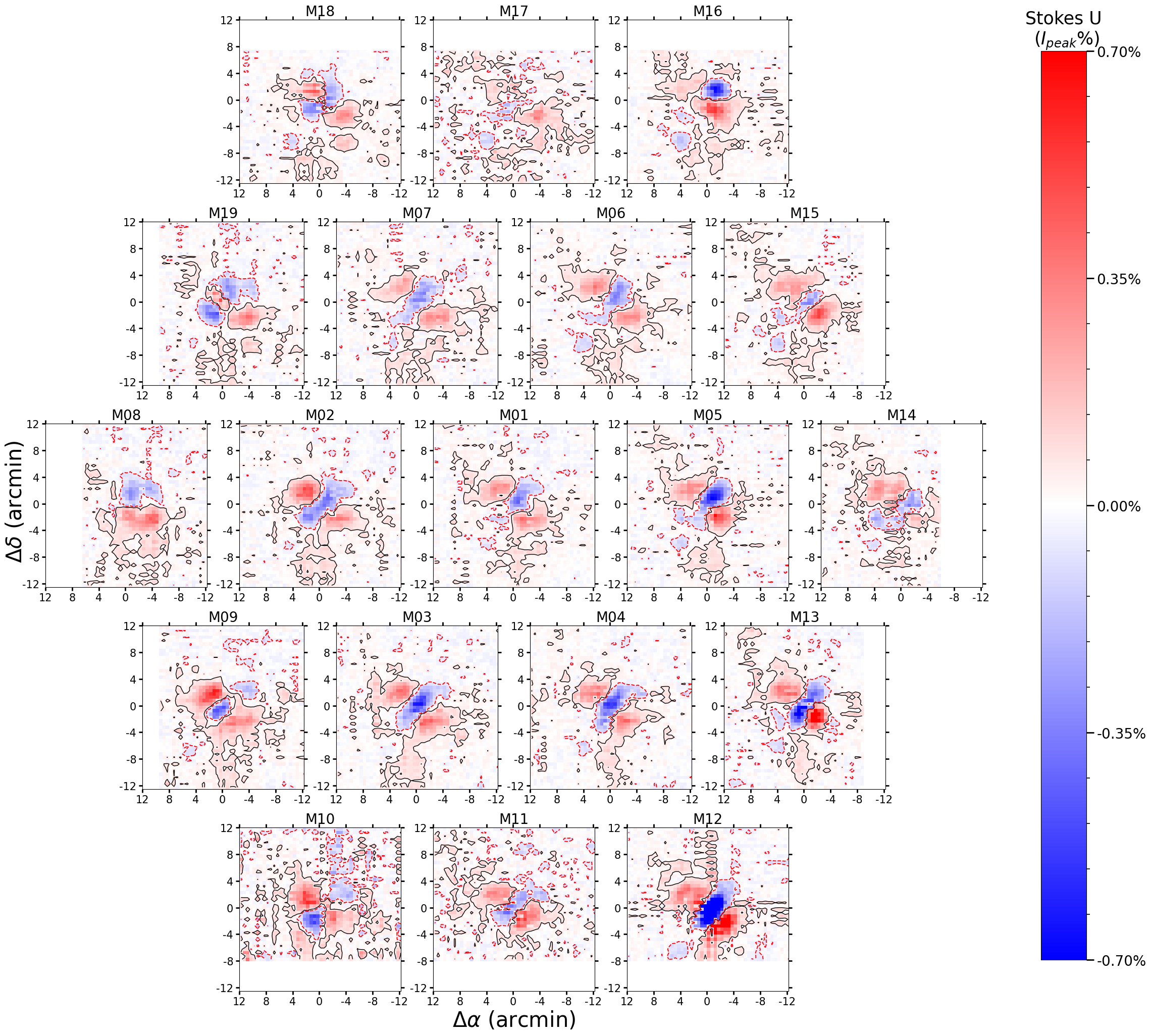}
\caption{Stokes $U$ mapping of J1407+2827 for the 19 beams near 1420 MHz. To compare with the simulated beam patterns, the mapping data have been mirror-symmetrized. Black solid contours represent $U/I_{peak}$ at 0.05\%, while red dashed contours represent $U/I_{peak}$ at $-0.05\%$.}
\label{fig14}
\end{figure*}

\begin{figure*}
\centering
\includegraphics[width=\textwidth, angle=0]{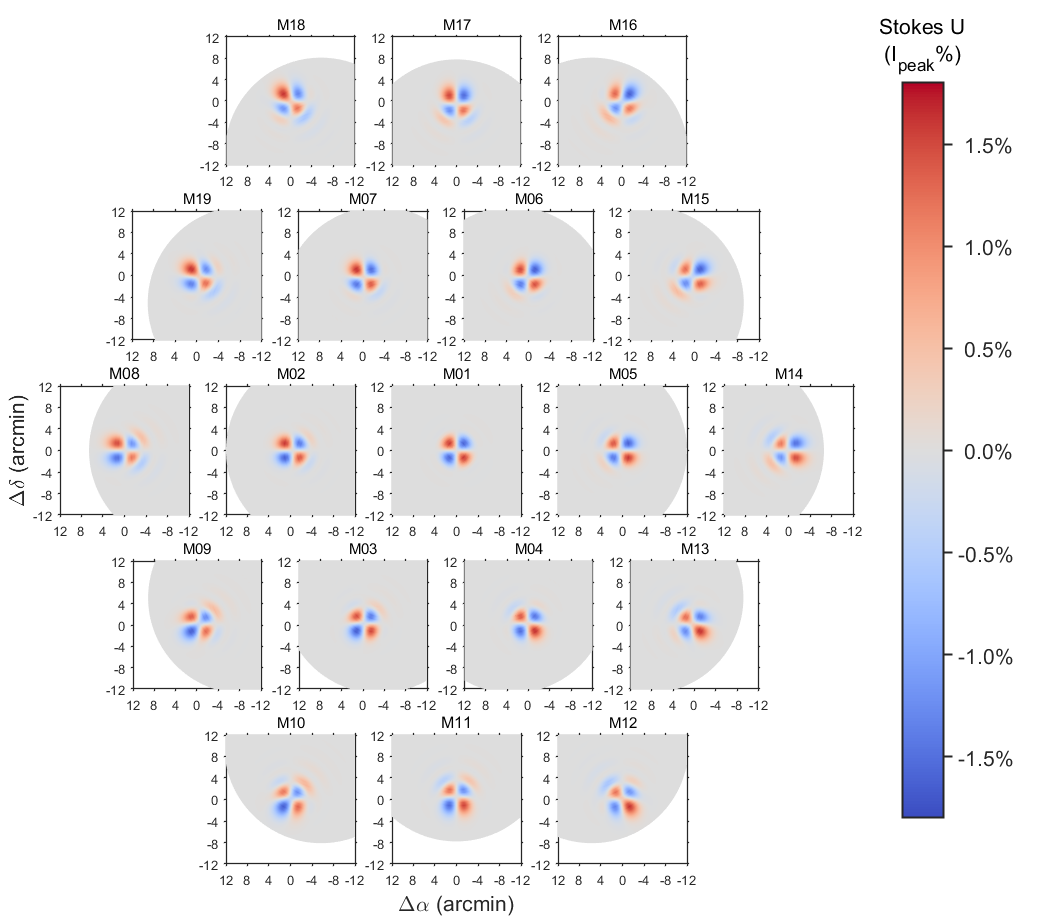}
\caption{Simulated Stokes $U$ patterns of all 19 individual beams.}
\label{fig15}
\end{figure*}

\begin{figure*}
\centering
\includegraphics[width=\textwidth, angle=0]{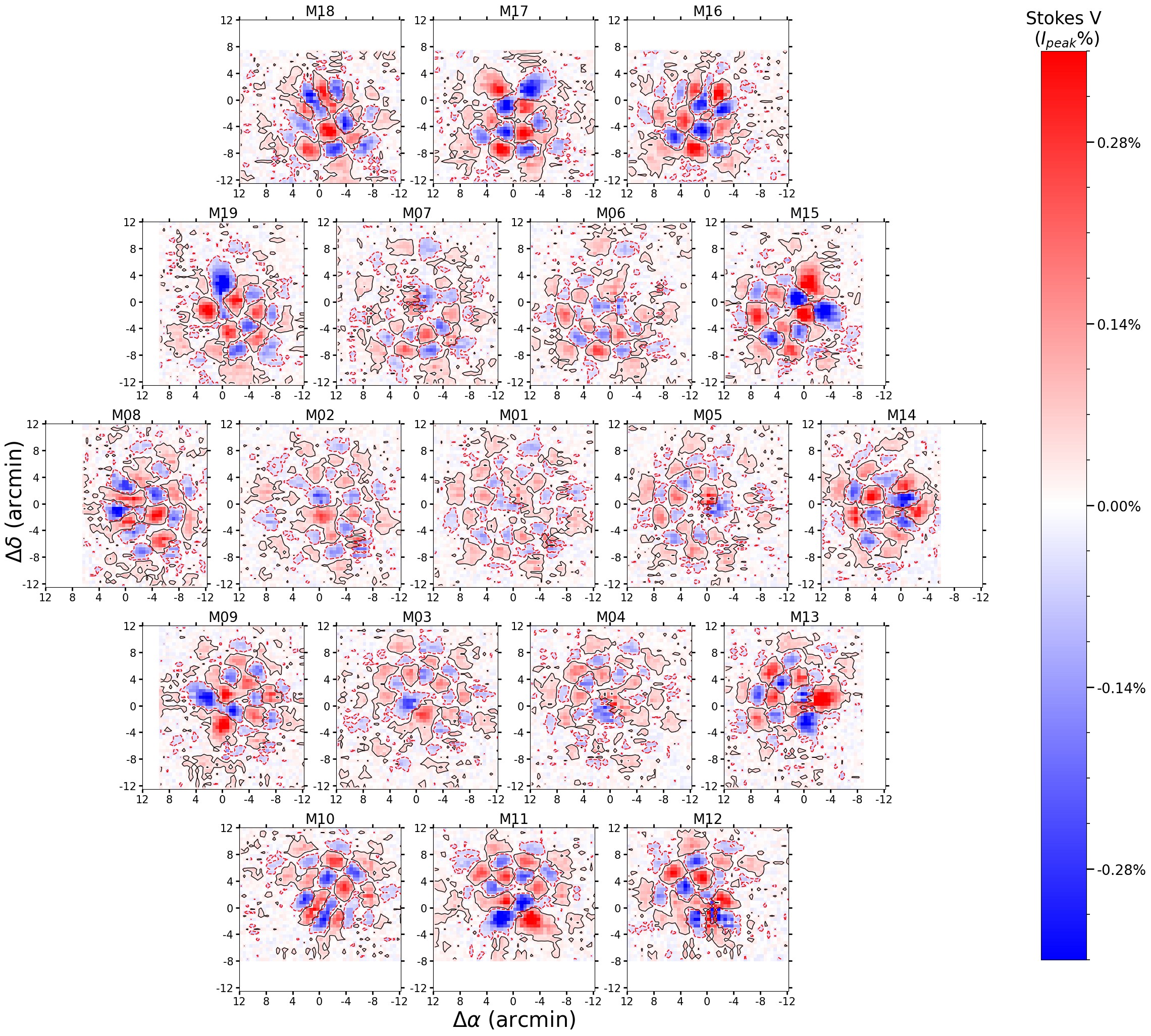}
\caption{Stokes $V$ mapping of J1407+2827 for the 19 beams near 1420 MHz. To compare with the simulated beam patterns, the mapping data have been mirror-symmetrized. Black solid contours represent $V/I_{peak}$ at 0.03\%, while red dashed contours represent $V/I_{peak}$ at $-0.03\%$.}
\label{fig16}
\end{figure*}

\begin{figure*}
\centering
\includegraphics[width=\textwidth, angle=0]{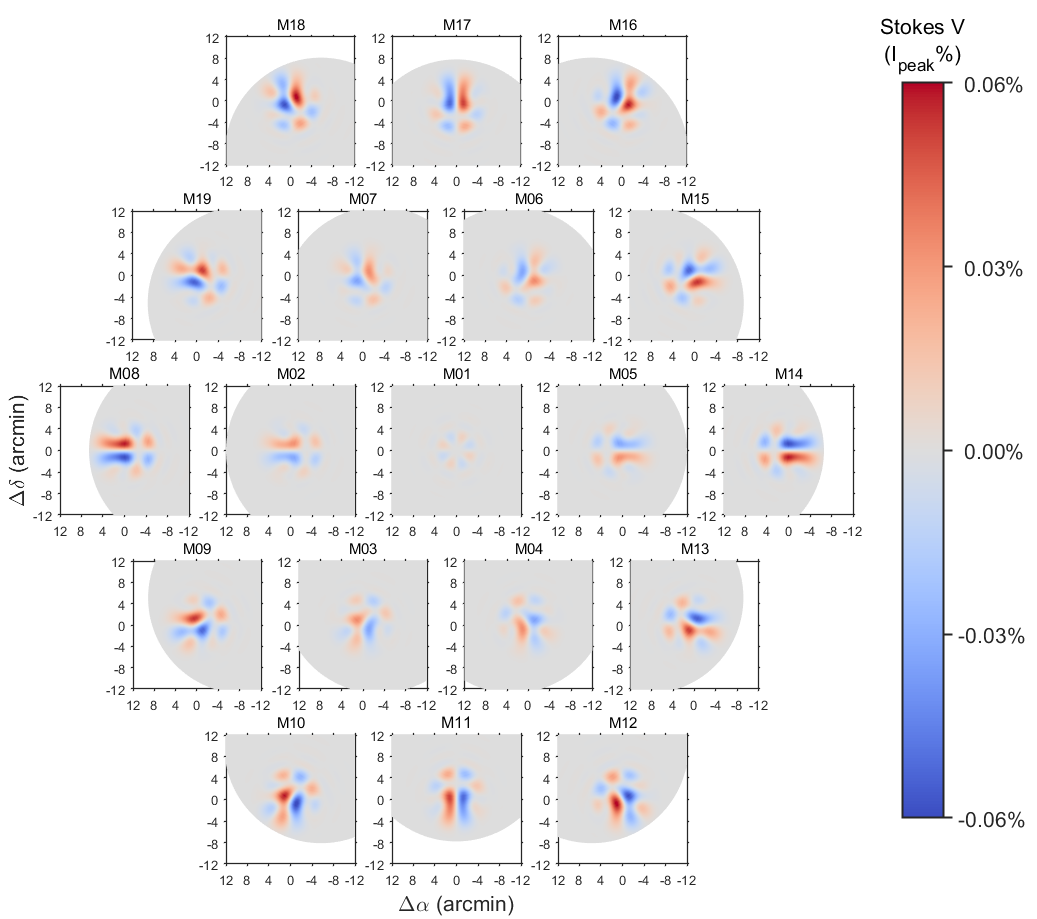}
\caption{Simulated Stokes $V$ patterns of all 19 individual beams.}
\label{fig17}
\end{figure*}

\begin{figure*}
\centering
\begin{minipage}{0.48\textwidth}
    \centering
    \includegraphics[width=\textwidth, angle=0]{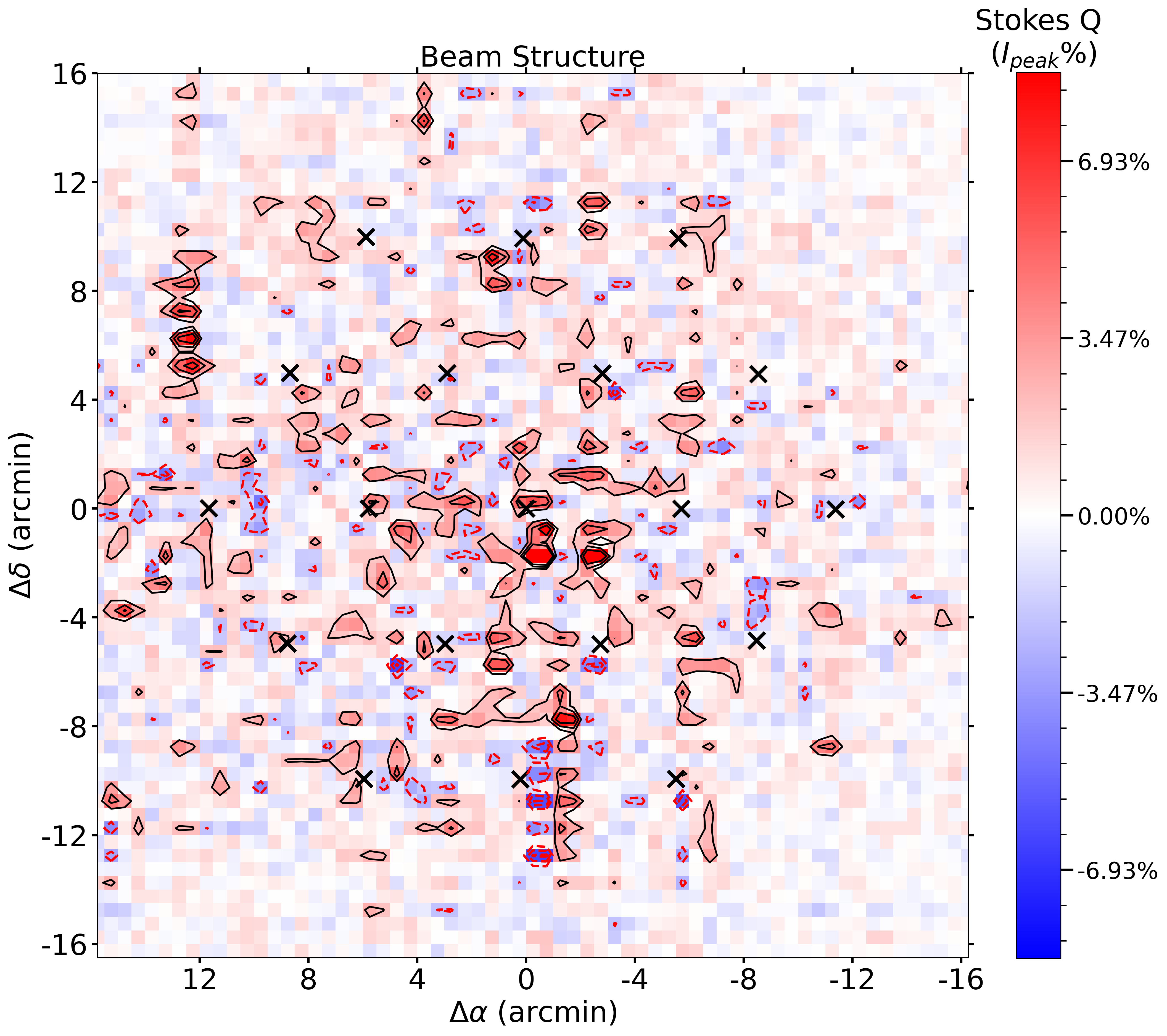}
\end{minipage}%
\hfill
\begin{minipage}{0.48\textwidth}
    \centering
    \includegraphics[trim=.7cm 0cm 1.2cm 0cm,clip,width=\textwidth, angle=0]{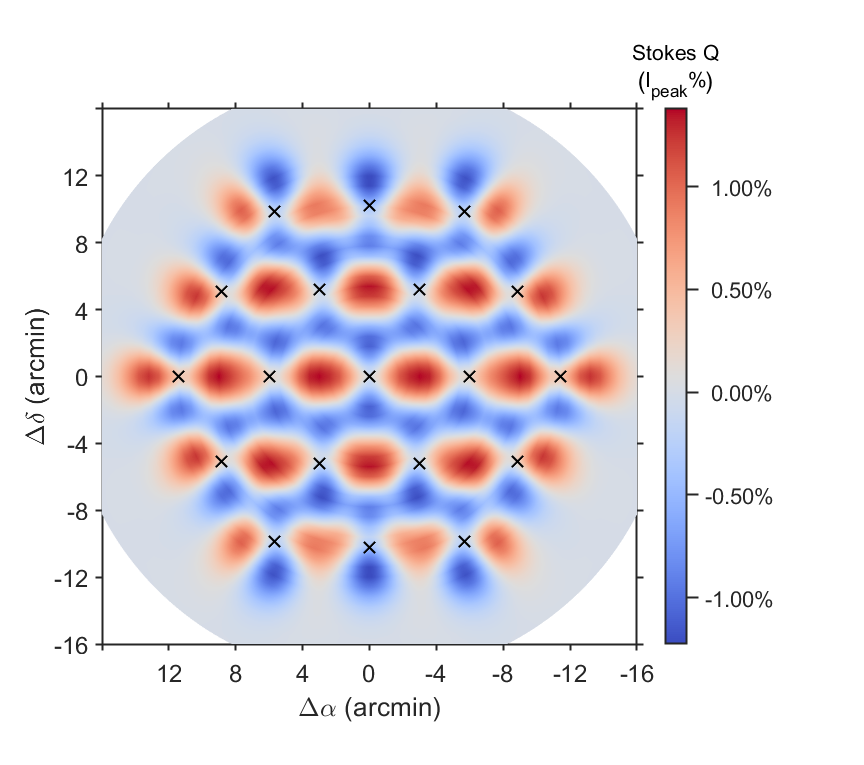}
\end{minipage}
\caption{\textbf{Left:} Combined response of the FAST 19-beam receiver for Stokes $Q$ towards J1407+2827. Black contours represent $Q/I_{peak}$ at 2\%, 4\%, and 6\%, respectively, while red dashed contours represent $Q/I_{peak}$ at $-2\%$, $-4\%$, and $-6\%$, respectively. \textbf{Right:} Simulated Stokes $Q$ patterns for the combined beam. The cross marks represent the 19 beam centers.}
\label{fig18}
\end{figure*}

\begin{figure*}
\centering
\begin{minipage}{0.48\textwidth}
    \centering
    \includegraphics[width=\textwidth, angle=0]{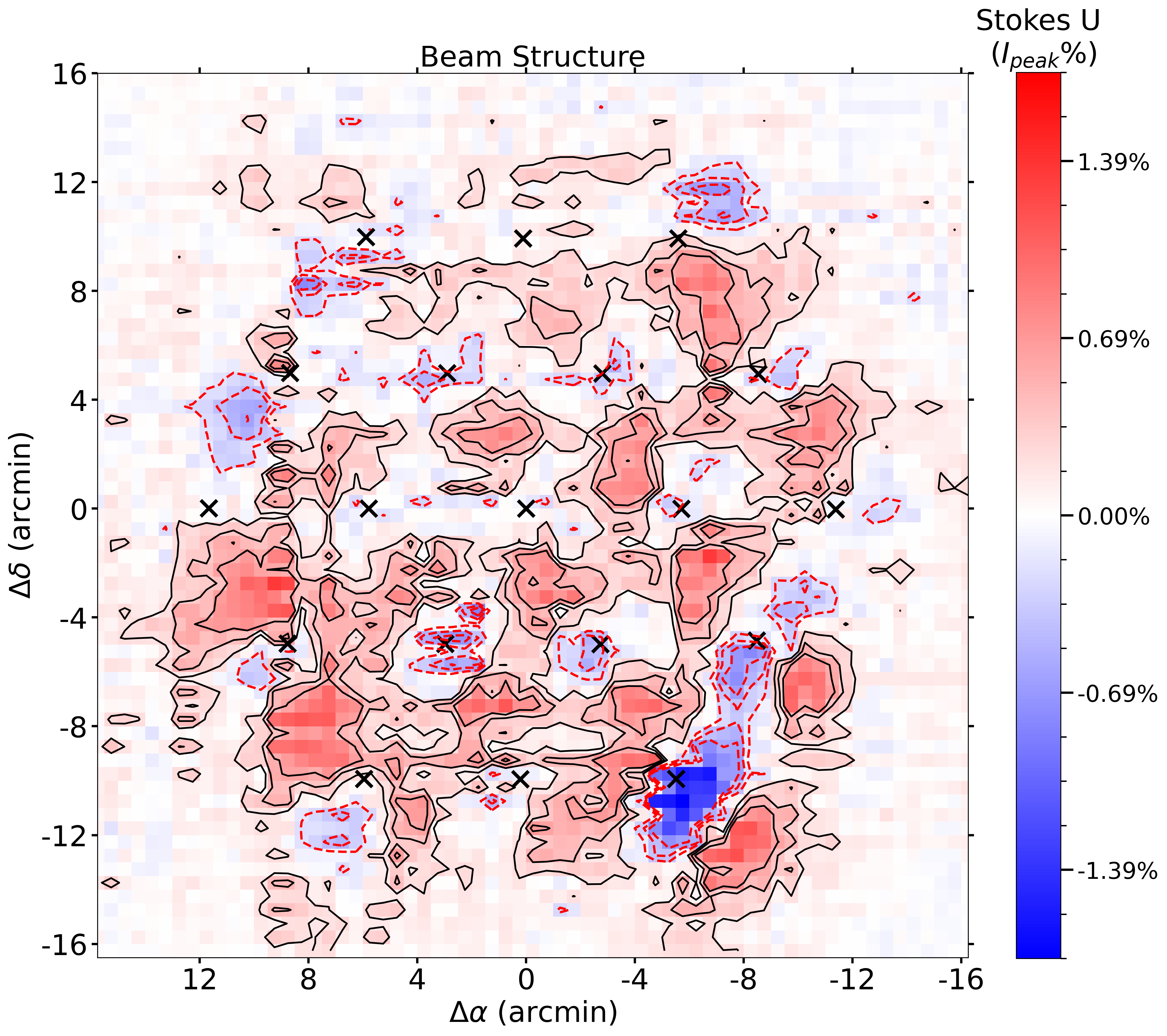}
\end{minipage}%
\hfill
\begin{minipage}{0.48\textwidth}
    \centering
    \includegraphics[trim=.7cm 0cm 1.2cm 0cm,clip,width=\textwidth, angle=0]{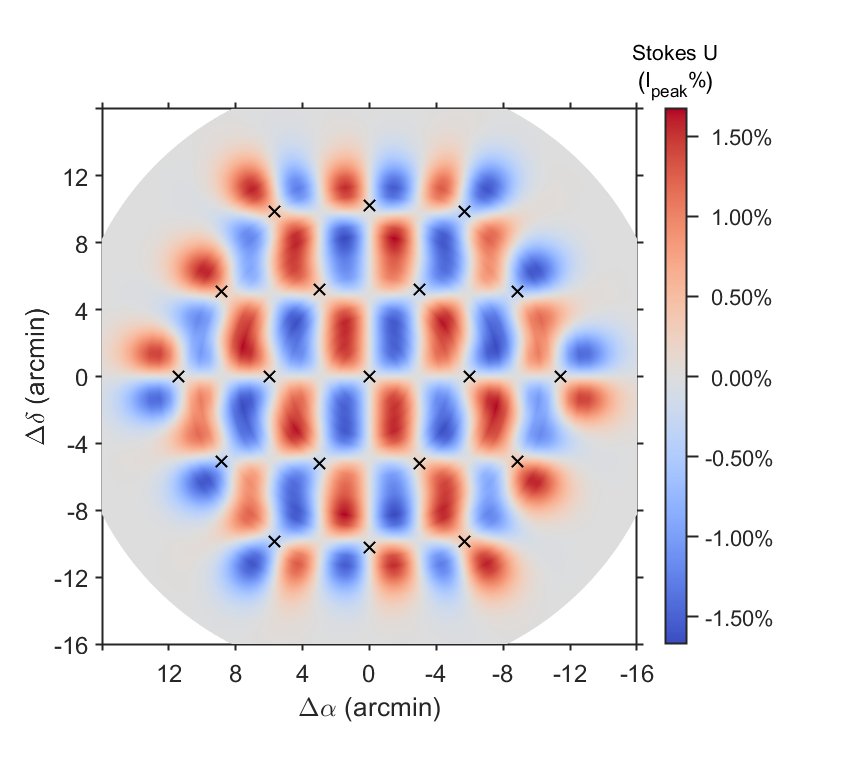}
\end{minipage}
\caption{Similar to Figure \ref{fig18}, but for Stokes $U$. Black contours represent $U/I_{peak}$ at 0.2\%, 0.4\%, and 0.6\%, respectively, while red dashed contours represent $U/I_{peak}$ at $-0.2\%$, $-0.4\%$, and $-0.6\%$, respectively.}
\label{fig19}
\end{figure*}

\begin{figure*}
\centering
\begin{minipage}{0.48\textwidth}
    \centering
    \includegraphics[width=\textwidth, angle=0]{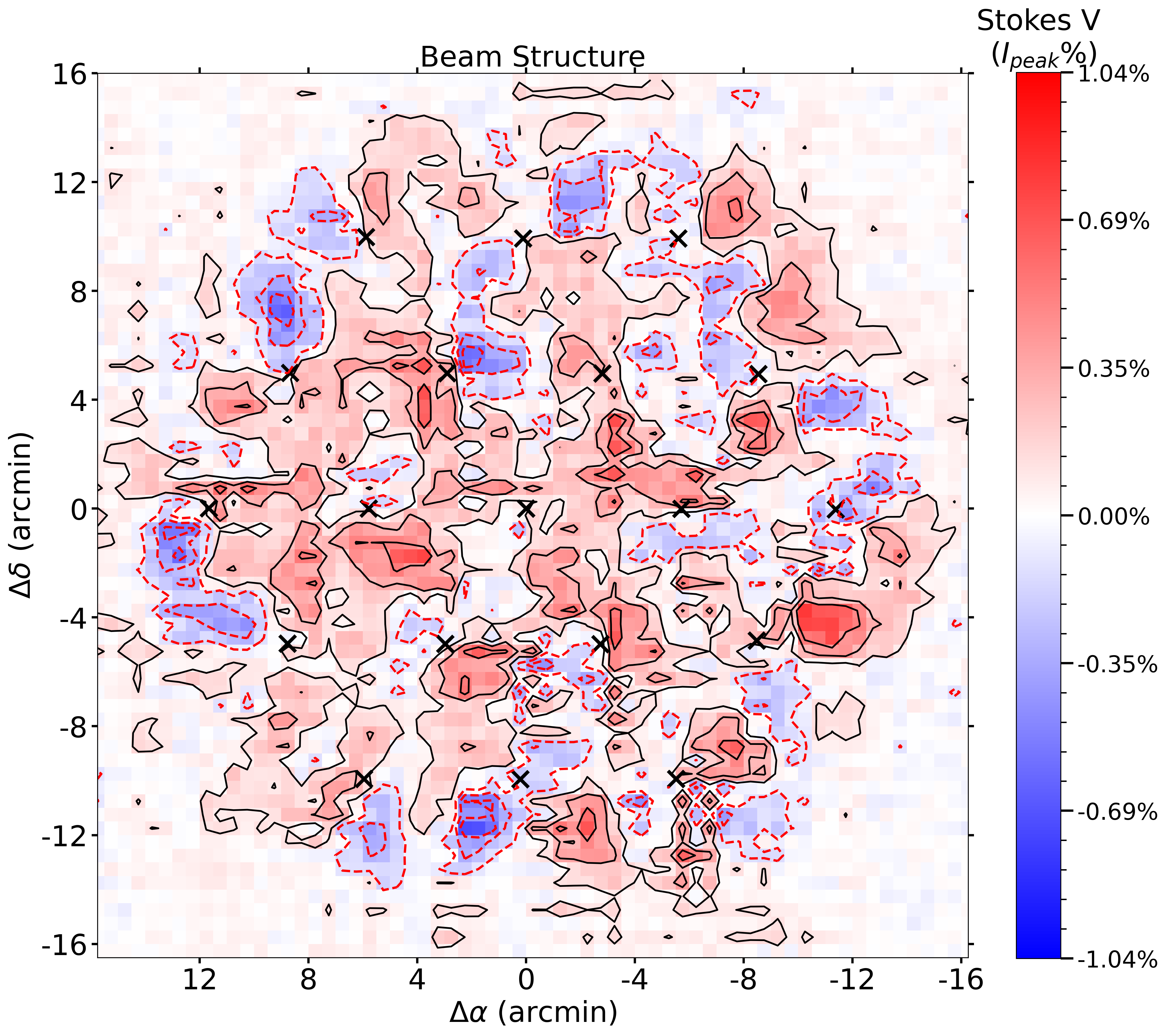}
\end{minipage}%
\hfill
\begin{minipage}{0.48\textwidth}
    \centering
    \includegraphics[trim=.7cm 0cm 1.2cm 0cm,clip,width=\textwidth, angle=0]{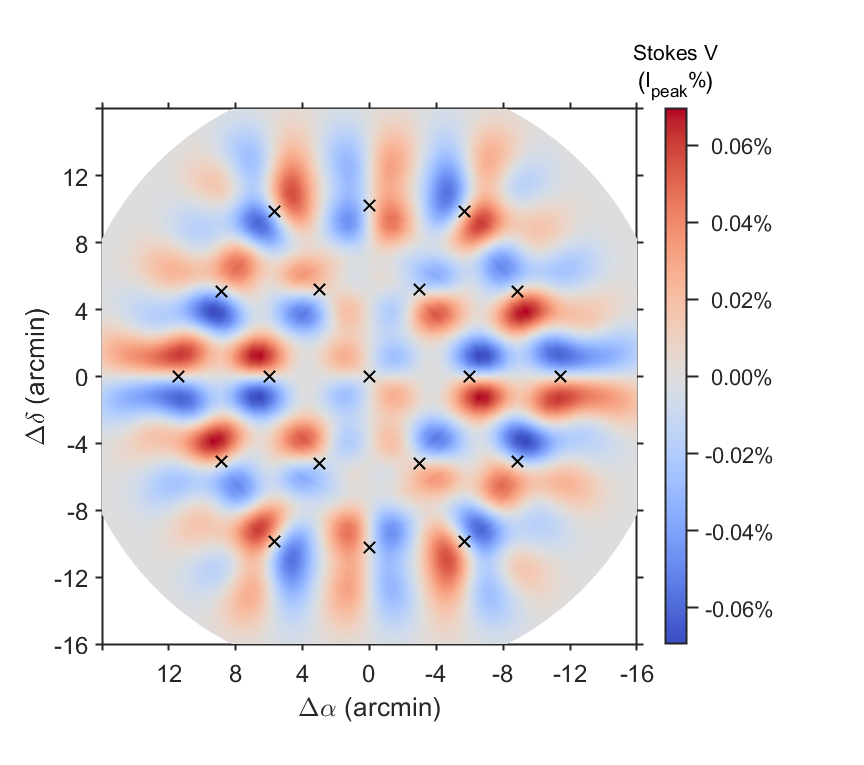}
\end{minipage}
\caption{Similar to Figure \ref{fig18}, but for Stokes $V$. Black contours represent $V/I_{peak}$ at 0.1\%, 0.3\%, and 0.5\%, respectively, while red dashed contours represent $V/I_{peak}$ at $-0.1\%$, $-0.3\%$, and $-0.5\%$, respectively.}
\label{fig20}
\end{figure*}

\subsection{Stokes $Q$, $U$, and $V$ beams}
The FAST receiver is equipped with 19 feeds designed for orthogonal linear polarizations, each followed by a cryogenic orthomode transducer (OMT), a temperature-controlled noise injection system, and low-noise amplifiers, which generate the $X$ and $Y$ signals for the two polarization channels \citep{2020Fan}. The measurement of beam shapes for polarized Stokes parameters entails two approaches: observing an unpolarized source or subtracting the polarization emission of a polarized source to emulate an unpolarized source. Unlike the response in Stokes $I$, the telescope's response in these parameters can be either positive or negative. J1407+2827 is a weakly polarized source\footnote{https://science.nrao.edu/facilities/vla/docs/manuals \allowbreak /obsguide/modes/pol}. The blazar J0854+2006 has been confirmed to be polarized in the Stokes $Q$ and $U$ parameters but not in Stokes $V$. These two sources can be utilized to measure the Stokes parameters $Q$, $U$, and $V$, which comprise three components: (1) the polarization of the source itself, (2) leakage from Stokes $I$, and (3) features in the Stokes $Q$, $U$, and $V$ beam patterns such as beam squint and beam squash~\citep{Robishaw2021}. Beam squash, which appears as a four-lobed cloverleaf pattern, is a linear polarization response that can arise from surface irregularities of the dish or the ellipticity of the feed radiation pattern~\citep{Robishaw2021}. In contrast, beam squint refers to the circular polarization response that occurs when a feed antenna is tilted or displaced from the focal point of a reflector. This effect is typically observed in the Stokes $V$ beam pattern and manifests as a two-lobed structure.

\begin{table}
  \centering
  \scriptsize
  \caption{Polarization measurements of J0854+2006 and J1407+2827.}
  \label{tab3}
  \renewcommand\tabcolsep{4.0pt} %
  \begin{tabular}{ccccc}
  \hline
   Source &Date	&Stokes $Q$&Stokes $U$&Stokes $V$\\
    &$(year)$ & $(Q/I_{peak})$& $(U/I_{peak})$& $(V/I_{peak})$ \\
  \hline
J0854+2006& 2020 &  3.04\%  &  4.66\% & $-0.17$\%\\
J0854+2006& 2021 &  1.15\%  &  2.42\% &  0.28\%\\
J1407+2827& 2022 &  1.95\%  & $-0.25$\% &  0.09\%\\
  \hline

  \end{tabular}
\end{table}

In our polarization calibrations, the Stokes parameters of J0854+2006 and J1407+2827 were measured, as shown in Table \ref{tab3}. Note that the Stokes Q measurements are mainly dominated by gain fluctuations. The Stokes U and V measurements below 1\% are dominated by instrumental polarization hence presenting upper limits for the intrinsic polarizations. In addition, the linear polarization has not been corrected with the ionospheric Faraday rotation, which is likely the origin of different Stokes Q and U values of J0854+2006 between 2020 and 2021. To remove the leakage from Stokes $I$ of J0854+2006 and J1407+2827, we used the percentage of the Stokes $I$ intensity profile to represent the leakage contribution. For the linearly polarized source J0854+2006, we further applied a Gaussian profile fit to model its intrinsic linear polarization. After subtracting these components, the residuals reveal the responses of beam structures in Stokes $Q$, $U$, and $V$. For example, Figure \ref{fig11} shows the beam structures for the central beam, with the colorbar being labeled with the ratio of the flux to the peak flux of Stokes $I$ (64.44 K, 67.50 K, and 28.85 K for data observed in 2020, 2021, and 2022, respectively) from the central beam, $I_{\rm peak}\%$. The simulated beam patterns for the central beam in Stokes $Q$, $U$, and $V$ are also depicted. In Stokes $Q$, all observations exhibit significantly higher noise, which is much stronger than the beam structure and obscures features such as beam squash. Since the noise is observed only in Stokes $Q$ and not in $U$ or $V$, we believe it is more likely intrinsic to the properties of linearly polarized receivers rather than solely attributed to radio frequency interference (RFI). Specifically, for linearly polarized receivers, Stokes $Q$ tends to be considerably noisier than $U$ and $V$ due to time-dependent gain fluctuations in the $E_X$ and $E_Y$ channels. As described in \citet{heiles2002} (Section 5.2), for a nonpolarized source, the error in Stokes $Q$ is directly proportional to gain fluctuations, while Stokes $U$ and $V$ remain independent of such errors. Thus, gain errors can introduce noise in $Q$ without affecting $U$ or $V$, because if $E_XE_Y$ = 0 (as it is for a nonpolarized source), the product remains zero even in the presence of gain fluctuations. In Stokes $U$, the beam squash is evident as a four-lobed cloverleaf pattern, where lobes on opposite sides of the beam center exhibit identical signs. The Stokes $U$ contours at $0.04\%$, which mark the squash structure, are roughly consistent between the J0854+2006 and J1407+2827 maps. Furthermore, these results align closely with the simulated beam squash pattern. Since beam squash is inherent to the feed pattern ellipticity, it is expected to be constant with time. 

Regarding Stokes $V$, all observations consistently reveal a more complex but weaker beam structure. The two-lobed beam squint structure is evident, accompanied with a secondary eight-lobed structure featuring an alternating arrangement of positive and negative lobes. These secondary structures align well with the predictions of the EM simulation. Theoretically, there should not be a squint $V$ pattern in the center beam, as shown in the simulation. The observed beam squint in the center beam suggests that the central horn is either laterally displaced away from the focus or tilted with respect to the focus-vertex line.  This can possibly occur from a combination of a few things: (1) the focus cabin is not placed exactly so that the center horn's phase center is at the focus of the vertex, so a slight positional inaccuracy will incur a squint; (2) the receiver as mounted in the focus cabin could have the plane that contains all the feed horns slightly tilted with respect to the horizontal axis of the focus cabin so that the normal to this receiver plane is not pointed towards the vertex; (3) the receiver is mounted inside the focus cabin so that the feed-horn-plane is aligned with the focus cabin's horizontal plane, but the focus cabin itself is slightly tilted, which in turn would cause the receiver plane's normal vector to point off of the vertex of the dish. This may also explain the nearly hundredfold difference in squint structure intensity between the observed results and the model predictions. Despite both J0854+2006 and J1407+2827 being unpolarized in Stokes $V$, the Stokes $V$ beam maps made using J0854+2006 have a higher noise than the map made using J1407+2827 owing to the data quality. For the central beam, the responses from beam squints remain consistently below $0.3\%$, with contours at $0.02\%$ delineating the squint structure. This demonstrates the robust reliability of the central beam in detecting Stokes $V$ polarization.

The Stokes $Q$, $U$ and $V$ mapping results of J1407+2827 for each individual beam together with the simulation results are depicted in Figures \ref{fig12}, \ref{fig13}, \ref{fig14}, \ref{fig15}, \ref{fig16}, and \ref{fig17}, respectively. As shown in these figures, Stokes $Q$ is much noisier, with no clear beam structure detected in any of the beams. Stokes $U$ exhibits a clear beam squash structure: for the central and inner beams, this structure remains consistent, with positive lobes on either side (in red) clearly separated, while the center region shows negative values. Consequently, the negative lobes on both sides (in blue) form an almost continuous shape, creating a unified structure. Similarly, the outer beams generally exhibit squash patterns, though the specific features vary across beams. For example, M10 shows a relatively clear squash pattern, while others, such as M11, share similarities with inner beams like M02 in certain aspects. M12, which is similar to M11, exhibits even stronger intensity. Overall, the $U/I_{peak}$ across all beams is below 0.7\%, with no significant intensity difference between inner and outer beams. The simulation predicts similar results, indicating that certain lobes in the outer beams’ squash structure are stretched, leading to a pattern distinct from that of the inner beams, while no intensity difference is observed between the inner and outer beams. In Stokes $V$, the two-lobed squint structure and secondary structures are clearly evident in all beams. Comparing with the central beam, inner beams exhibit a more concentrated distribution in one direction that correlates strongly with the position angle. This effect is even more pronounced in the outer beams, where the structures are more compact, resulting in a significant increase in intensity due to the horns being displaced further from the focus. The beam squint orientation is distinctly observed in both the inner and outer beams, with the two-lobed squint structure aligning with the angle of each beam’s position angle. Overall, the $V/I_{peak}$ across all beams remains below 0.4\%, with the central beam demonstrating the best performance. These results align closely with the beam squint predictions by the simulation, particularly concerning the stretching of the lobes. For instance, the model of M08 clearly illustrates the positive and negative lobes being stretched into a horizontal configuration, which is in excellent agreement with the observed results. Additionally, the model also indicates that the intensity of the squint structure in the outer beams is significantly greater than that in the inner beams.

Figures \ref{fig18}, \ref{fig19}, and \ref{fig20} depict the combined response of FAST in the Stokes $Q$, $U$, and $V$ parameters for J1407+2827 together with the theoretical model. Intrinsic noise in Stokes $Q$, likely due to gain fluctuations in linearly polarized receivers, obscures the beam squash. Conversely, the performance of all beams in Stokes $U$ exhibits favorable characteristics, with responses from beam squashes having $U/I_{peak} \leq 1.5\%$. For Stokes $V$, it is observed that the combined response of all 19 beams displays beam squint intensity having $V/I_{peak} \leq 1\%$, indicating the highest level of performance among the Stokes $Q$, $U$, and $V$ parameters. Due to data quality issues, the beam squash and squint structures exhibited in the combined response of the 19 beams differ somewhat from the results of the theoretical model; however, both show alternating positive and negative structures in Stokes $U$ and $V$. Furthermore, numerous observed vertical and horizontal structures are present, indicating that these features are not artifacts of the data but rather the result of overlapping squash and squint structures from different beams. These findings affirm the exceptional capabilities of FAST in detecting the Stokes $U$ and $V$ parameters, providing a robust foundation for conducting in-depth investigations into the interstellar magnetic field, particularly concerning Zeeman multi-beam observations. We recognize that instrumental polarization and contamination effects, particularly near strong Stokes $I$ sources, may influence fractional polarization measurements in some cases. However, these effects are unlikely to impact the detection of the Zeeman effect, which relies on spectral differences in RCP and LCP components rather than diffuse polarized emission. Ongoing studies are investigating how instrumentally polarized beam patterns interact with total intensity profiles to produce artificial signals in Stokes $V$, particularly in cases where velocity gradients align with the beam squint. Insights from these efforts, including potential first-order corrections, could guide future improvements in FAST’s capabilities for polarization studies.

\section{Conclusions}\label{con}
This study provides a comprehensive analysis to characterize the beam structures of the Five-hundred-meter Aperture Spherical radio Telescope (FAST) across the Stokes $I$, $Q$, $U$, and $V$ parameters, encompassing all 19 beams. We use a multi-beam on-the-fly (MultiBeamOTF) mapping pattern designed for simultaneous sky surveys using these 19 beams, directed towards polarization calibrators J1407+2827 and J0854+2006. The theoretical model was also developed, with the electromagnetic simulation packages CST and GRASP-10 used to compute the telescope’s complete radiation patterns across all Stokes parameters. Our investigation confirms the symmetry of the central beam and the asymmetry of the off-center beams. The central beam of FAST shows a symmetrical pattern without detectable sidelobes, which is significant as it indicates a well-behaved central beam, crucial for accurate observations. The off-center beams exhibit significant asymmetrical shapes; these shapes were modeled using log-normal and Gaussian methodologies. This asymmetry can affect the accuracy of data and needs to be accounted for in analyses.

Focusing on the Stokes $I$ parameter, we report fitting results revealing variations in $\sigma_1$, relative beam efficiency, $\phi_{\rm dif}$, beam size, position angle, and $P_{\rm slb}/P_{\rm beam}$. Our analysis concludes that inner beams demonstrate superior beam efficiency and beam size compared to outer beams. This suggests that the inner beams are more effective in collecting data and have better performance. The distribution of position angles demonstrates a consistent angular spacing of 29 degrees between the outer beams and 60 degrees between the inner beams, which is a direct result of the hexagonal layout of the feedhorns. Quantitative examination reveals that the sidelobe responses of the inner beams contribute approximately 2\% of the main lobe flux, increasing to 5\% for outer beams, with a peak at 6.8\% of M16. This indicates that sidelobe contamination is relatively minor in the inner beams but more pronounced in outer beams. These results can provide valuable support for work related to data within the narrow frequency range centered at 1420.4 MHz.

Furthermore, we study the responses of FAST's 19-beam receiver across the Stokes $Q$, $U$, and $V$ parameters. The analysis of Stokes $U$ and $V$ reveals distinct beam squash and squint structures across the 19 beams. Stokes $U$ shows a clear four-lobed cloverleaf squash structure, particularly in the central and inner beams, where positive lobes are well-defined on either side, and negative values dominate the center region. Although the overall $U/I_{peak}$ across all beams remains below 0.7\% without significant differences between inner and outer beams. The theoretical model supports these observations, indicating that certain lobes in the outer beams are stretched, resulting in a unique pattern distinct from the inner beams.

In Stokes $V$, the two-lobed squint structure and a secondary eight-lobed structure featuring an alternating arrangement of positive and negative lobes are evident in all beams. Compared with the central beam, inner beams exhibit a more concentrated distribution in one direction that correlates strongly with the position angle. This trend is even more pronounced in the outer beams, where compact structures lead to increased intensity due to the horns’ displacement from the focus. The beam squint orientation aligns with the respective position angles, and the overall $V/I_{peak}$ remains below 0.4\%, with the central beam performing the best. These findings align closely with the beam squint predictions by the simulation.

\clearpage
\begin{acknowledgements}
This work was supported by the National Natural Science Foundation of China (Grants No.12403037 and No.12373026), the Leading Innovation and Entrepreneurship Team of Zhejiang Province of China (Grant No.2023R01008), and the Key R\&D Program of Zhejiang, China (Grant No.2024SSYS0012).

\end{acknowledgements}



\clearpage

\label{lastpage}

\end{document}